\def\lessim{\lower.5ex\hbox{$\; \buildrel < \over \sim \;$}}
\def\simgt{\lower.5ex\hbox{$\; \buildrel > \over \sim \;$}}

\newcommand\aj{AJ}
\newcommand\apj{ApJ}
\newcommand\apjl{ApJ}
\newcommand\apjs{ApJ}
\newcommand\araa{ARA\&A}
\newcommand\mnras{MNRAS}
\newcommand\pasp{PASP}

\newcommand\aap{A\&A}

\documentclass{cupconf}
\usepackage{natbib}
\usepackage{graphicx}

  \checkfont{eurm10}
  \iffontfound
    \IfFileExists{upmath.sty}
      {\typeout{^^JFound AMS Euler Roman fonts on the system,
                   using the 'upmath' package.^^J}%
       \usepackage{upmath}}
      {\typeout{^^JFound AMS Euler Roman fonts on the system, but you
                   don't seem to have the}%
       \typeout{'upmath' package installed. CUPconf.cls can take advantage
                 of these fonts,^^Jif you use 'upmath' package.^^J}%
       \providecommand\mathup[1]{##1}%
       \providecommand\mathbup[1]{##1}%
      }
  \else
    \providecommand\mathup[1]{#1}%
    \providecommand\mathbup[1]{#1}%
  \fi

  \checkfont{msam10}
  \iffontfound
    \IfFileExists{amssymb.sty}
      {\typeout{^^JFound AMS Symbol fonts on the system, using the
                'amssymb' package.^^J}%
       \usepackage{amssymb}%
         
         \let\geq=\geqslant
      }{}
  \fi

  \IfFileExists{amsbsy.sty}
    {\typeout{^^JFound the 'amsbsy' package on the system, using it.^^J}%
     \usepackage{amsbsy}}
    {\providecommand\boldsymbol[1]{\mbox{\boldmath $##1$}}}

\title[Black Holes in Deep Surveys]{Supermassive Black Holes in Deep Multiwavelength Surveys}

\author[C. M. Urry and E. Treister]{C.\ns M\ls E\ls G\ls A\ls N\ns U\ls R\ls R\ls Y$^1$ 
\and E\ls Z\ls E\ls Q\ls U\ls I\ls E\ls L\ns T\ls R\ls E\ls I\ls S\ls T\ls E\ls R$^2$}

\affiliation{$^1$Department of Physics, Yale University, P.O. Box 208121, New Haven, CT 06520-8121, USA\\[\affilskip]
	$^2$European Southern Observatory, Casilla 19001, Santiago 19 Chile }

\begin{document}

\maketitle

\begin{abstract}
In recent years deep X-ray and infrared surveys have provided an efficient way 
to find accreting supermassive black holes,
otherwise known as active galactic nuclei (AGN), 
in the young universe.
Such surveys can, unlike optical surveys, find AGN obscured by 
high column densities of gas and dust. 
In those cases, deep optical data show only the host galaxy, 
which can then be studied in greater detail than in unobscured AGN. 
Some years ago the hard spectrum of the X-ray ``background" 
suggested that most AGN were obscured. 
Now GOODS, MUSYC, COSMOS and other surveys have confirmed this 
picture and given important quantitative constraints on AGN demographics. 
Specifically, we show that most AGN are obscured at all redshifts 
and the amount of obscuration depends on both luminosity and redshift, 
at least out to redshift $z\sim2$, 
the epoch of substantial black holes and galaxy growth. 
Larger-area deep infrared and hard X-ray surveys will be 
needed to reach higher redshifts 
and to probe fully the co-evolution of galaxies and black holes.
\end{abstract}

\firstsection

\section{Cosmic Growth of Black Holes and Galaxies}

Abundant evidence indicates that the growth of a 
supermassive black hole is closely tied to 
the formation and evolution of the surrounding galaxy. 
The energy released from accretion onto the black hole 
affects star formation in the galaxy, 
probably limiting growth at the high- and low-mass ends, 
and of course the distribution and angular momentum of matter in
the galaxy governs the amount of matter accumulated by the black hole
\citep{silk98,king05,rovilos07}. 
Emergent energy from accretion is also a factor in understanding
ionization and radiation backgrounds \citep{hasinger00,lawrence01}. 
Understanding the growth history of these black holes is 
therefore critical to understanding the
global evolution of structure in the Universe.

Yet the demographics of supermassive black holes remain elusive.
The largest samples of quasars and Active Galactic Nuclei (AGN),
by which we mean supermassive black holes in a high accretion-rate 
phase\footnote{Some make a distinction between AGN and quasars, 
with the latter being above an arbitrary luminosity, typically $M_B=-23$~mag. 
Here we use the term AGN to refer to an actively accreting 
supermassive black hole above or below this luminosity.},
have been found through optical selection 
(e.g., the Sloan Digital Sky Survey quasar sample; 
\citealp{schneider02,schneider07}), 
but, at least locally, these are not representative 
of the larger AGN population.
Instead we need surveys less biased against obscured AGN.

There are three reasons to suspect that most AGN are obscured 
by large column densities of gas and dust. 
First, a large body of evidence suggests that local AGN have geometries
that are not spherically symmetric, and that different aspect angles present 
markedly different observed characteristics; this is referred to as AGN unification \citep{antonucci93,urry95}. 
Second, AGN are more common at high redshift ($z\sim2-3$),
where the average star formation rate is higher and
thus it is even more likely that gas and dust surround the 
galaxy nucleus than at $z\sim0$. 
Third, and most important, obscured AGN are required to explain the shape of
the X-ray ``background" radiation.

\begin{figure}
\begin{center}
\includegraphics[width=.53\textwidth]{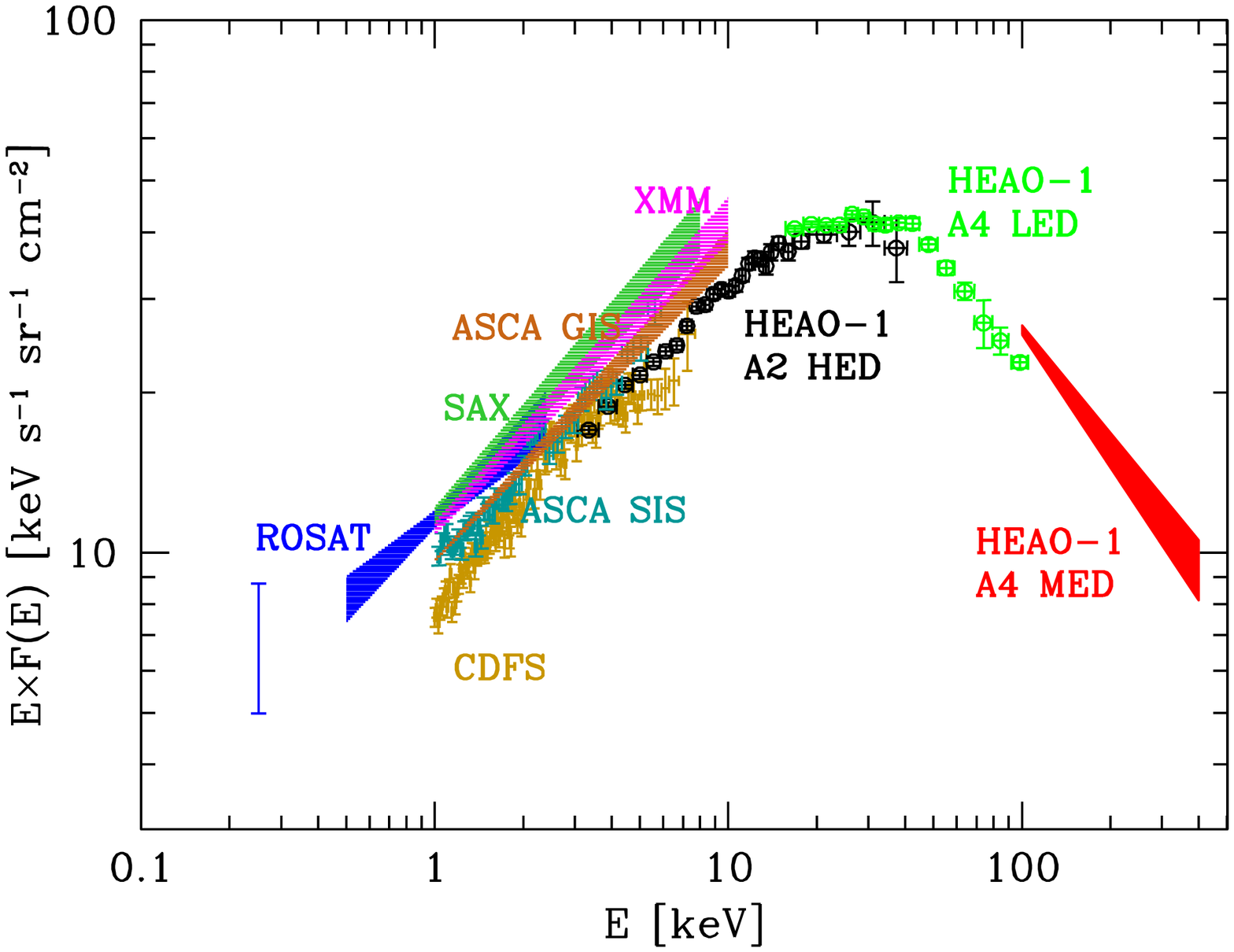}
\includegraphics[width=.43\textwidth]{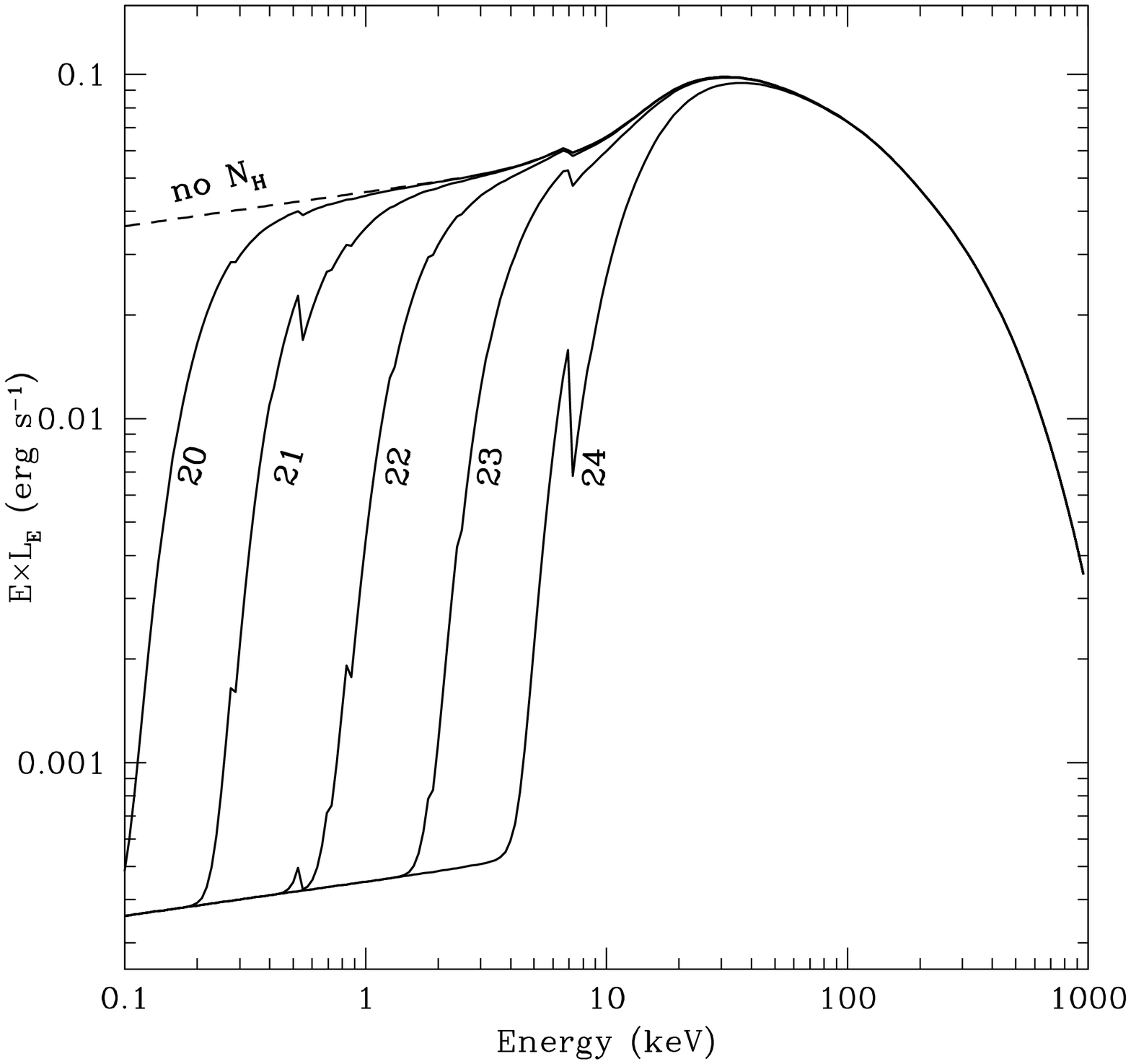}
\caption{
({\it Left}) The X-ray background spectrum 
(shown in units of energy per logarithmic band) 
is very hard, peaking near 30~keV, 
in contrast to the spectra of unobscured AGN which are
roughly horizontal in these units.
(Figure from \citealp{gilli04}.) 
({\it Right}) Large column densities of gas absorb 
low-energy photons via the photoelectric effect, 
such that emission from an obscured AGN peaks at 
an energy that increases with column density. 
Each line represents a factor of 10 increase in 
equivalent $N_H$ 
(lines are marked with $\log N_H$ in atoms/cm$^2$; 
assumes solar abundances and the cross
sections of \citealp{morrison83}). }
\label{fig_xrbgspec}
\end{center}
\end{figure}

The X-ray ``background" is actually the superposition of 
individual AGN that were not resolved in early X-ray experiments 
(hence the designation ``background"). 
As shown in Figure~1a, its spectrum peaks at $\sim30$~keV 
(i.e., this is where most of the energy is produced), 
much harder than the typical spectrum (roughly flat in these units) 
of an unobscured AGN. 
In obscured AGN, however, the softest X-ray photons have been 
absorbed via the photoelectric effect, and Compton-thick AGN 
(those with $N_H\geq 10^{24}$~cm$^{-2}$) 
actually peak at roughly 30~keV (Fig.~\ref{fig_xrbgspec}b).
Thus X-ray observations have long
indicated there is a large population of heavily obscured AGN
\citep{setti89,comastri95,gilli01}. 
While much of the X-ray background has been resolved at low
energies, recent work on X-ray deep fields suggests the hardest 
(most obscured) X-ray sources have yet to be detected 
(e.g., \citealp{worsley05}).

Relatively unbiased AGN samples require joint hard X-ray and infrared surveys.
Hard X-rays ($E>2$~keV) penetrate all but the thickest column densities
and efficiently locate black hole accretion.
The absorbed optical through soft X-ray emission heats the surrounding 
dust and is re-radiated in the infrared, at wavelengths that depend on the
dust temperature.
Meanwhile, high-resolution optical surveys (e.g., with HST) allow separation
of nuclear (AGN) and host galaxy light.
Today, NASA's three operating Great Observatories --- the Chandra X-ray Observatory,
the Spitzer Space Telescope, and the Hubble Space Telescope ---
enable matched X-ray, infrared, and optical surveys at 
unprecedented depth and resolution.

\section{Deep Multiwavelength Surveys}

The announcement in spring 2000 of the first
Spitzer Legacy opportunity led to 
the Great Observatories Origins Deep Survey (GOODS), 
to date the deepest wide-area multiwavelength survey carried out with 
Spitzer, Hubble, and Chandra \citep{dickinson03,giavalisco04}. 
By targeting the Spitzer Legacy, and later the Hubble Treasury, 
observations on the pre-existing Chandra deep fields, 
we leveraged substantial investments of observing time, 
probed AGN demographics at the peak of quasar activity ($z\sim1-2$), 
and enabled a wide range of other science,
described in the {\it Astrophysical Journal Letters} 
special edition of January 10, 
2004.\footnote{For more details see http://www.stsci.edu/science/goods.}

The GOODS data are deep enough to detect AGN 
to very high redshift ($z\simgt 6$), 
and the volume sampled ensures sizable AGN samples out to $z\sim3$. 
To sample larger volumes and to search effectively for 
rare objects like AGN or massive galaxies at high redshift, 
in 2002 we designed the MUSYC survey 
(Multiwavelength Survey by Yale and Chile), 
covering one square degree in two equatorial and two southern fields. 
Three of the four regions had already been observed extensively, 
including the Extended Chandra Deep Field South (ECDF-S) 
and the Extended Hubble Deep Field South (EHDF-S), 
thus leveraging substantial investments with Chandra, 
XMM-Newton, HST, and the largest ground-based 
telescopes.\footnote{The four MUSYC fields include the ECDF-S, 
for which substantial additional Chandra observations were 
later acquired by N. Brandt \citep{lehmer05,virani06}; 
the EHDF-S, for which there not yet any X-ray data; 
a field centered on a $z=6.3$ quasar at 1030+05, 
which has relatively deep XMM data; 
and a new field centered at 1256+01 (``Castander's Window"), 
an equatorial region with low 100-micron dust emission. 
The first two have now been imaged with Spitzer (IRAC and MIPS) 
and HST (ACS and NICMOS), and all fields are accessible with ALMA. 
For more details see http://www.astro.yale.edu/MUSYC
and \citet{gawiser06a,quadri07,gawiser06b,vandokkum06}.} 
Soon after starting MUSYC we helped start the 
COSMOS survey \citep{scoville07a}, 
a 2-square degree field centered at 1000+02 that has now been imaged 
extensively with HST, Spitzer, XMM, and Chandra.\footnote{COSMOS 
is now one of the largest and best covered deep multiwavelength survey fields, 
representing the investment of thousands of hours of major telescope time
\citep{scoville07b,sanders07,hasinger07,capak07,lilly07}.
For more details, see http://cosmos.astro.caltech.edu.}
The unprecedented combination of area and depth allow a 
wide variety of science, described in the September 2007 
special edition of the Astrophysical Journal Supplement. 
Other relevant multiwavelength surveys include 
the Lockman Hole \citep{hasinger01}, 
CLASXS \citep{yang04}, the Extended Groth Strip \citep{davis07}, 
SWIRE \citep{lonsdale03}, XBOOTES \citep{hickox06a}, 
HELLAS2XMM \citep{baldi02}, SEXSI \citep{harrison03}, 
CYDER \citep{treister05a}, CHAMP \citep{kim04}, and AMSS \citep{akiyama03}. 

\section{Finding Obscured AGN}

The question we set out to answer with GOODS, MUSYC, and COSMOS was, 
``Is there a substantial population of obscured AGN 
that is missed by traditional optical surveys?"
To answer this question we need to sample the AGN population at $z\sim1-2$, 
at the peak of the number density. 
The GOODS survey, and the Chandra Deep Field South (CDF-S) 
X-ray survey on which GOODS piggy-backed \citep{giacconi02},
were designed to sample the AGN population at $z\sim0.5-2$,
which includes the AGN that make up the X-ray background. 
Luminous AGN at higher redshifts could certainly be detected 
but the volume surveyed is too small to expect to see 
a reasonable number of them.

Early results in the CDFS and other deep X-ray fields 
showed there was a population of optically faint hard X-ray sources 
\citep{alexander01,franceschini02a,franceschini02b,mainieri05b}, 
which collectively comprised a large fraction of the integrated 
X-ray background intensity. 
However, as optical counterparts were identified and spectra obtained, 
several apparent problems emerged. 
First, the redshift distribution peaked at relatively low values, 
$z\lessim1$, lower than expected from the 
early population synthesis models for the X-ray background. 
Second, the fraction of X-ray sources that were identified as obscured, 
either because of high $N_H$ or absence of broad lines, 
was less than the canonical 3/4 seen at low redshift, 
and appeared to decline with redshift rather than increase, 
as specified by the best population synthesis model
at that time \citep{gilli01}. 
A number of authors pointed out these problems 
(e.g., \citealp{mainieri05b,rosati02,brandt05}), 
and \citet{franceschini02b} suggested that 
``the unification scheme based on a simple orientation effect 
fails at high redshifts" and that the production of the 
X-ray background by a collection of obscured AGN ``requires major revision." 

At the same time, these faint red X-ray AGN were copious emitters 
of infrared radiation \citep{treister04, treister06a},
fully consistent with the unification paradigm.
It also became apparent early on that, 
as for optically-selected quasars, 
the evolution of X-ray selected AGN is luminosity dependent, 
with high-luminosity AGN evolving earlier and more rapidly 
than low-luminosity AGN \citep{ueda03,cowie03}; 
this luminosity-dependent density evolution had not been incorporated 
in earlier population synthesis models for the X-ray background.
Finally, we suspected that selection effects could 
play a significant role in affecting the redshifts 
and optical identifications of survey sources.

Accordingly, we developed a comprehensive quantitative approach, 
based on the unification scenario and 
incorporating the most recent, best luminosity function and evolution,
to predict the number counts and redshift distributions 
at any wavelength (for the moment, infrared through X-ray, 
though it could be generalized) for surveys of 
arbitrary area, depth, and wavelength. 
Here we describe the quantitative interpretation of the 
multiwavelength data from GOODS, MUSYC, COSMOS and 
other deep multiwavelength surveys, 
including the dependence of the obscured fraction of AGN 
on luminosity and redshift. 
Specifically, we show that most AGN are obscured at all redshifts; 
that the fraction of obscured AGN decreases with luminosity; 
and that it increases with redshift, at least out to redshift $z\sim2.5$, 
an epoch of substantial black hole and galaxy growth. 
Deep infrared and hard X-ray surveys over larger areas 
will be needed to reach higher redshifts 
and to probe fully the co-evolution of galaxies and black holes.

\subsection{Connecting X-Ray, Optical, and Infrared Surveys}

Our approach was to connect surveys at different wavelengths 
by assuming something sensible about AGN spectral energy distributions (SEDs), 
then combining those with well-measured luminosity functions and evolution 
to understand the source counts and redshift distributions of AGN selected 
at a given flux limit at any wavelength 
(\citealp{treister04,treister06a,treister06b}; 
analogous approaches have been taken by \citealp{ballantyne06} 
and \citealp{dwelly06}). 
Simultaneously, we constrain the same AGN population 
to fit the X-ray background \citep{treister05b}. 
An alternative approach is to model only the X-ray spectra of AGN 
and to fit the X-ray background alone with a mixture of obscured 
and unobscured AGN \citep{comastri95,gilli01,gilli07}; 
this constrains the demographics but does not connect 
the X-ray sources to those detected at optical and infrared wavelengths 
--- and thus does not use those additional constraints on the AGN demographics,
nor does it allow a quantitative estimate of the important effect of 
optical or infrared flux limits on the survey content or 
spectroscopic identifications. 

Briefly, our procedure was as follows: 
We started with the underlying AGN demographics, 
described by an AGN luminosity function that incorporates 
dependence on absorbing column density, 
and the luminosity-dependent evolution of this function \citep{ueda03}. 
Because this luminosity function is based on hard X-ray observations, 
it is relatively free of bias against obscured AGN.

\begin{figure}
\begin{center}
\includegraphics[width=.45\textwidth]{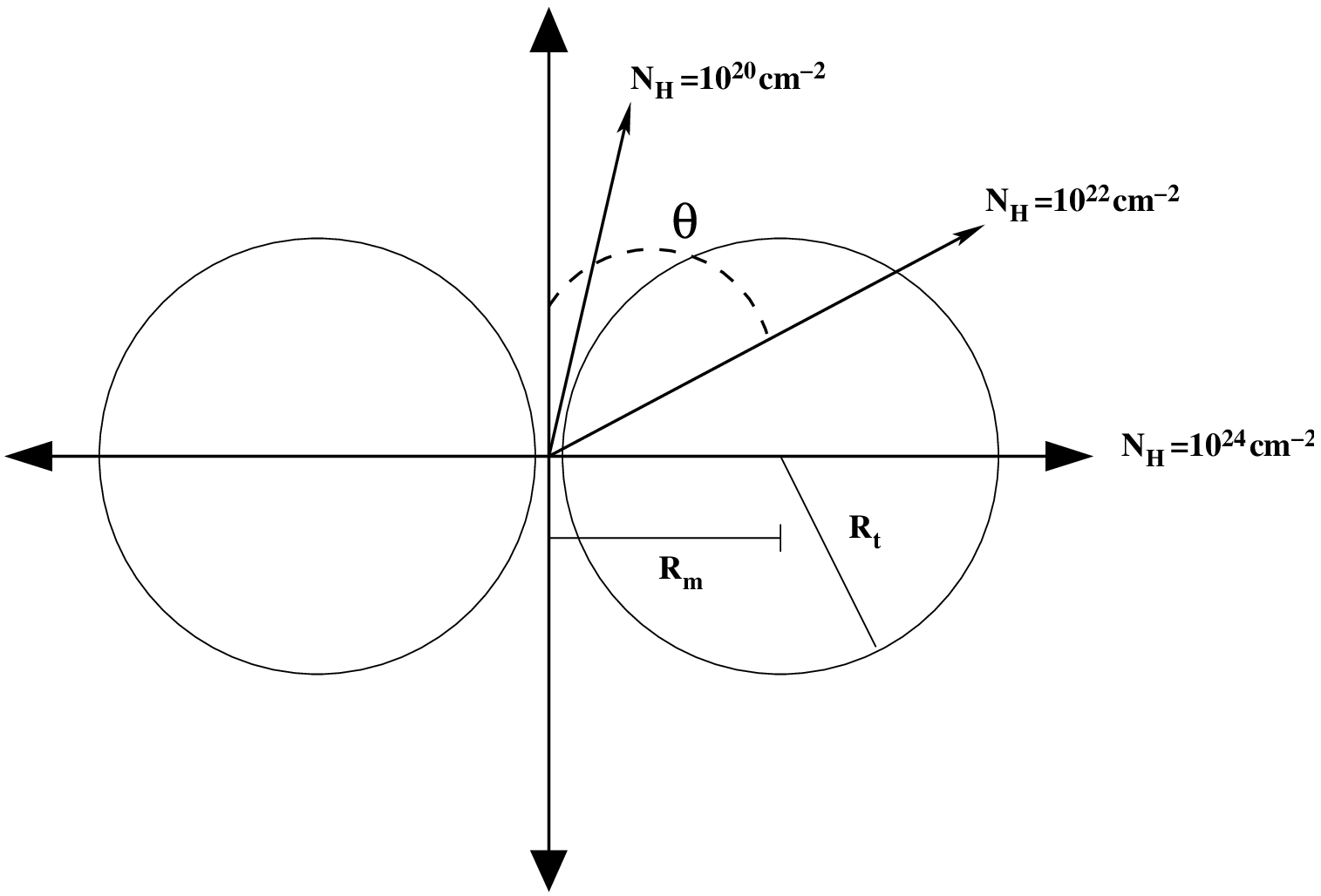}
\includegraphics[width=.45\textwidth]{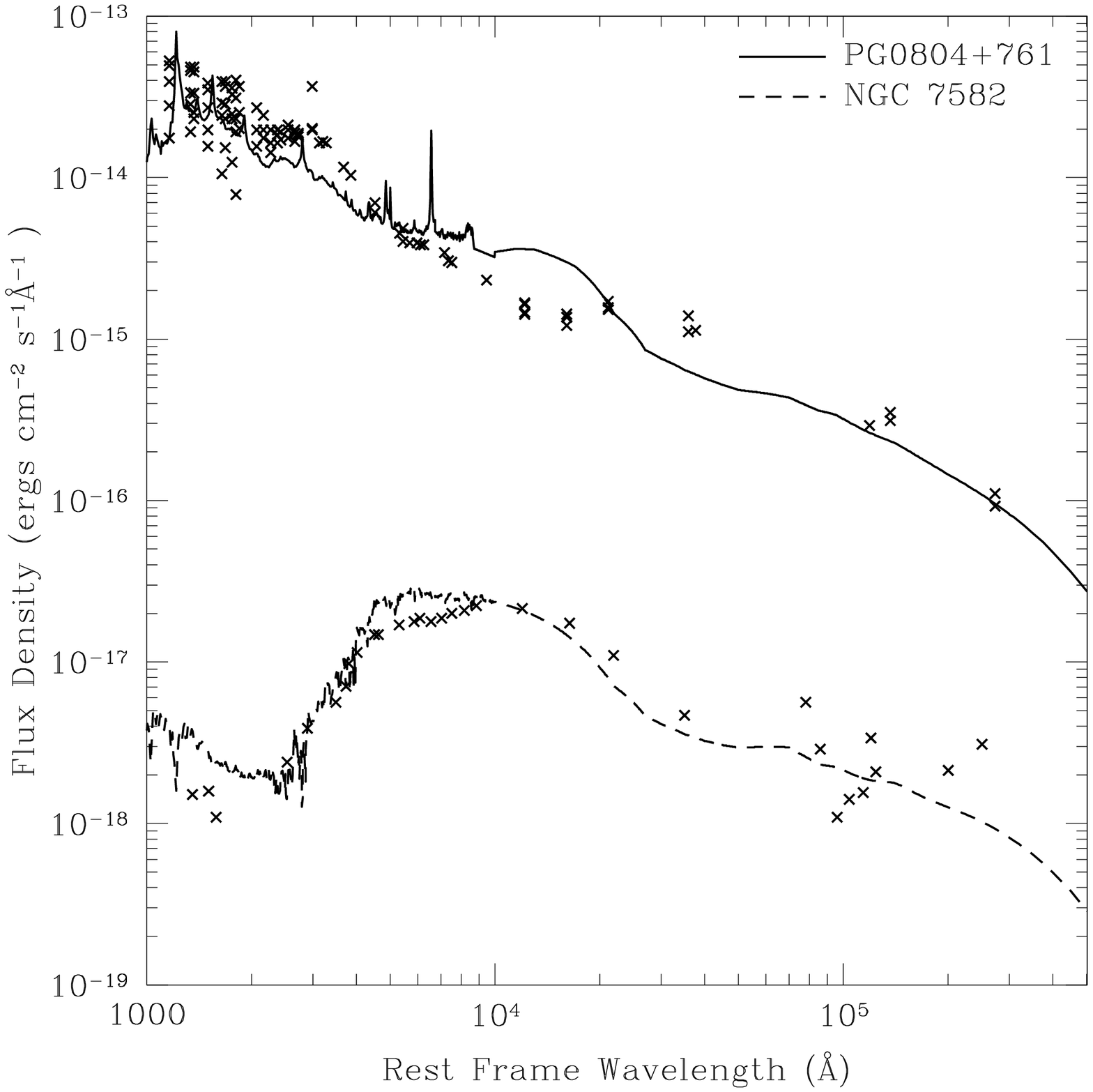}
\caption{
The infrared spectra of AGN were modeled by dust emission 
\citep{nenkova02} 
from a clumpy torus geometry ({\it left}) that, 
assuming random orientations and adjusting 
the torus geometry to give a 3:1 ratio of column densities
above:below $N_H=10^{22}$~cm$^{-2}$,
yields an $N_H$ distribution consistent with various observational estimates
(see text for details). 
({\it Right}) 
Two examples of model SEDs ({\it lines}), 
which are fully determined from $L_X$ and $N_H$.
These fit very well the observed SEDs of local unobscured 
({\it datapoints, top}) and
obscured ({\it bottom}) AGN, with no free parameters.
The composite model SEDs include infrared dust emission,
reddened quasar spectra (keyed to $N_H$ value),
and an $L_*$ host galaxy, linked to the X-rays by the known
optical-to-X-ray ratio (which depends on $L_X$).
}
 \label{figSED}
\end{center}
\end{figure}

We then developed a set of SEDs, based on the unification paradigm, 
that represent AGN with a wide range in intrinsic luminosity and 
absorbing column density (parameterized in terms of the 
neutral hydrogen column density, $N_H$, along the line of sight). 
Specifically, at optical wavelengths ($\lambda$=0.1-1 microns), 
we used a Sloan Digital Sky Survey composite quasar spectrum 
\citep{vandenberk01} plus Milky-Way-type reddening laws 
and a standard dust-to-gas ratio to convert $N_H$ to $A_V$; 
we also added an $L_*$ elliptical host galaxy, 
which is the dominant component for heavily obscured AGN. 
In the X-ray ($E>$0.5 keV), we assumed a power-law spectrum 
with photon index $\Gamma = 1.9$, typical of unobscured AGN, 
absorbed by column densities in the range
$\log N_H = 20$-24~cm$^{-2}$. 
To describe the infrared part of the SEDs ($\lambda>$1 micron), 
which in the unification paradigm includes radiation from 
dust heated by absorbed ultraviolet through soft X-ray photons, 
we used dust emission models by \citet{nenkova02} in a 
clumpy torus geometry (Fig.~\ref{figSED}a), 
converting to $N_H$ from viewing angle assuming random orientations. 
The resulting $N_H$ distribution is completely consistent 
with various observational estimates 
\citep{ueda03, dwelly06, risaliti99, comastri95, tozzi06, gilli07}. 
AGN models with the same intrinsic X-ray luminosities 
were normalized at 100 microns. 
To connect the ultraviolet and X-ray parts of the SED 
we used the standard dependence of X-ray to optical luminosity ratio 
on luminosity (e.g., \citealp{steffen06}).

The SED models, parameterized in terms of $L_X$ 
(as a proxy for intrinsic luminosity) and 
$N_H$ (which depends only on viewing angle and torus geometry), 
describe extremely well the local population of AGN. 
There is some freedom in the choice of torus geometry, of course; 
we selected an aspect ratio that would produce three times as many AGN 
with $N_H$ greater than $10^{22}$~cm$^{-2}$ compared to 
smaller column densities (Fig.~\ref{figSED}a), 
which is roughly the observed local ratio \citep{risaliti99}. 
Figure~\ref{figSED}b shows the observed SEDs of two local AGN, 
one unobscured (top) and one heavily obscured (bottom), 
compared to the model SED corresponding to the observed $L_X$ and $N_H$. 
With no free parameters, the fit is remarkably good overall, 
meaning that this representation accurately reflects the 
distribution of photons across a very wide wavelength range.

Combining these model SEDs with the AGN luminosity function 
and evolution yields a distribution of sources as a 
function of spatial location and wavelength. 
For a given wavelength and flux limit, then, 
our calculation produces the expected number counts and 
redshift distribution of AGN. 
This population can be filtered by multiple flux limits 
(for example, one could look at the X-ray counts and redshift distribution 
of sources above arbitrary flux limits in the X-ray, optical, 
and infrared bands) and/or by other properties (e.g., $N_H$ value). 
Our procedure was to generate the expected content of real surveys, 
and to compare to observations, in order to constrain 
the underlying AGN demographics, 
specifically, the AGN luminosity function, 
its evolution, its dependence on $N_H$, 
and the family of AGN SEDs (which includes the torus properties).

\begin{figure}
\begin{center}
\includegraphics[width=.45\textwidth]{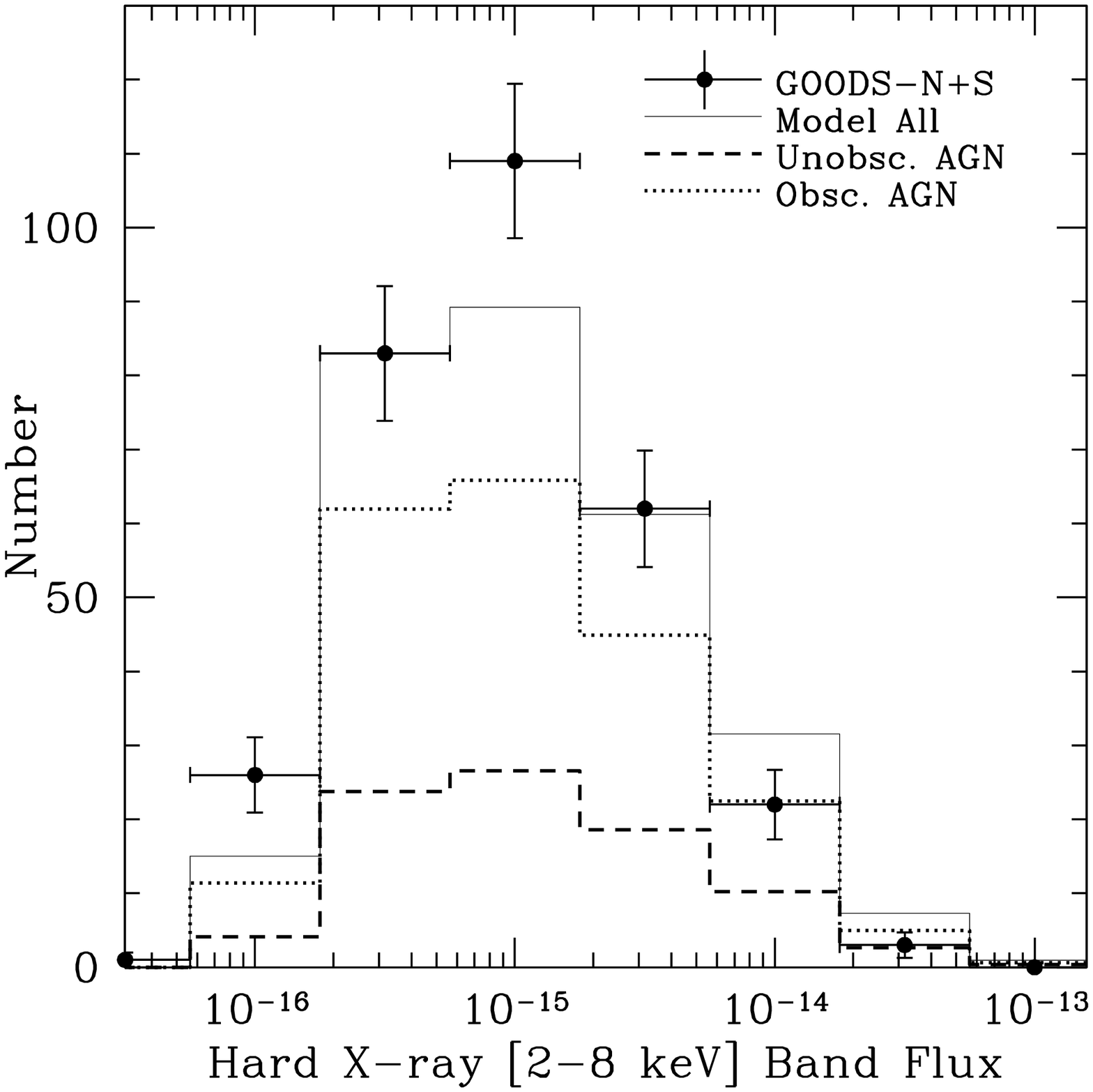}
\includegraphics[width=.45\textwidth]{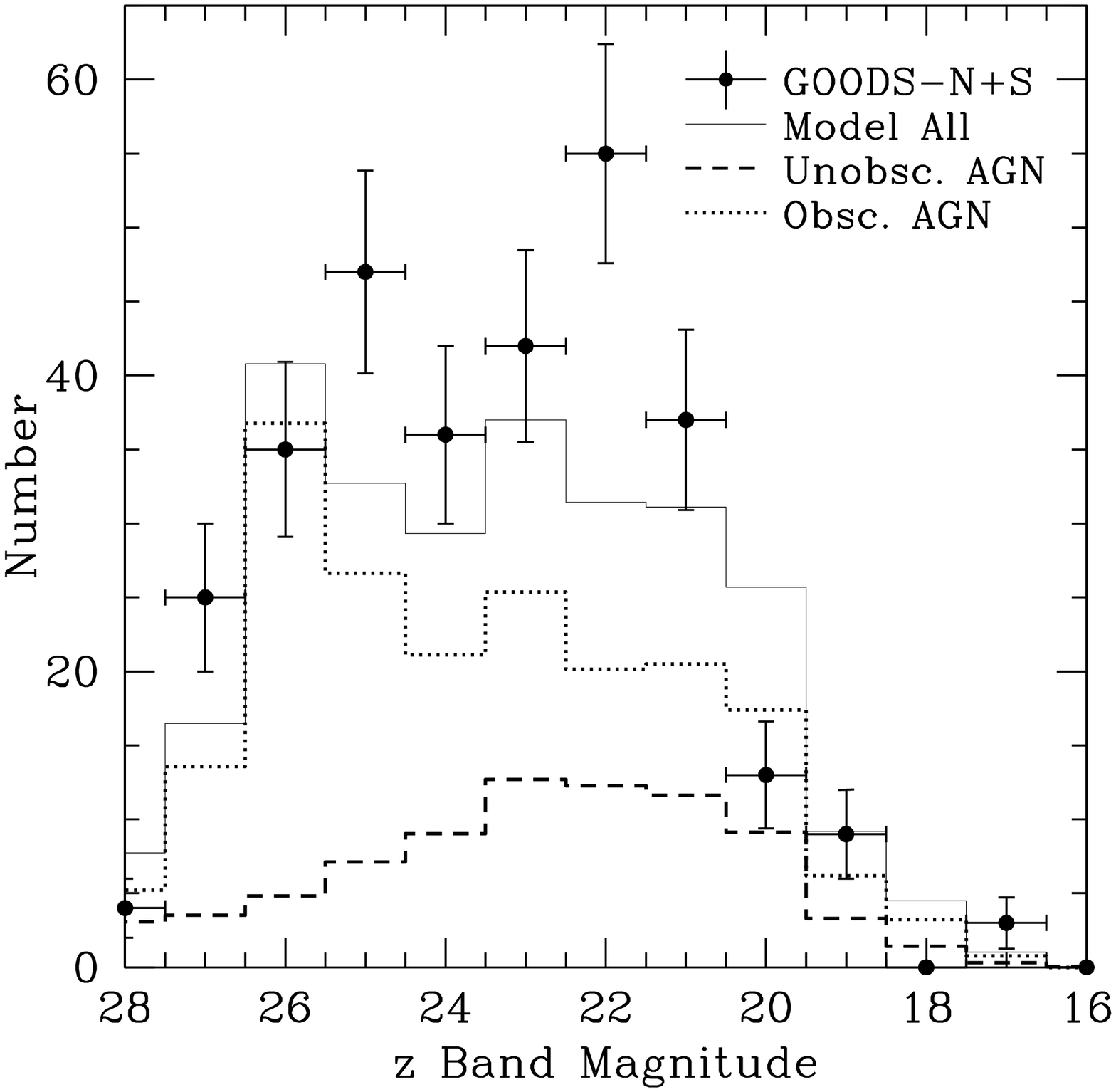}
\caption{
X-ray ({\it left}) and optical ({\it right}) counts observed in
the GOODS fields ({\it data points}) agree well with a simple
unification model ({\it solid line}), even with no dependence
of torus geometry on luminosity or redshift \citep{treister04}.
Inclusion of the luminosity dependence (found later in larger
AGN samples) gives essentially the same result for the GOODS survey.
Unobscured AGN ($N_H < 10^{22}$~cm$^{-2}$)
dominate at bright optical magnitudes ({\it dashed line}) 
and obscured AGN ($N_H > 10^{22}$~cm$^{-2}$)
at faint z-band magnitudes ({\it dotted line}). 
For $R>24$~mag, the approximate limit for obtaining
decent optical spectral identifications with 8-meter class telescopes,
the vast majority of sources are obscured AGN.}
  \label{figCounts}
\end{center}
\end{figure}

\subsection{Quantitative Results from the Population Synthesis Model:
Number Counts, Redshift Distributions, and the X-Ray Background}
\label{secResults}

In a series of papers we showed that this straightforward model 
matches very well the observed properties of AGN samples. 
The first paper, \citet{treister04}, assumed for the sake of simplicity 
that the torus geometry (and hence the $N_H$ distribution) 
did not depend on luminosity or redshift. 
Even this simplest population synthesis model fits
beautifully the observed X-ray and optical counts in 
the deep GOODS survey (Fig.~\ref{figCounts}).
The source population at faint optical magnitudes, 
particularly the AGN too faint for identification with optical spectroscopy,
is dominated by obscured AGN.

\begin{figure}
\begin{center}
\includegraphics[width=.45\textwidth]{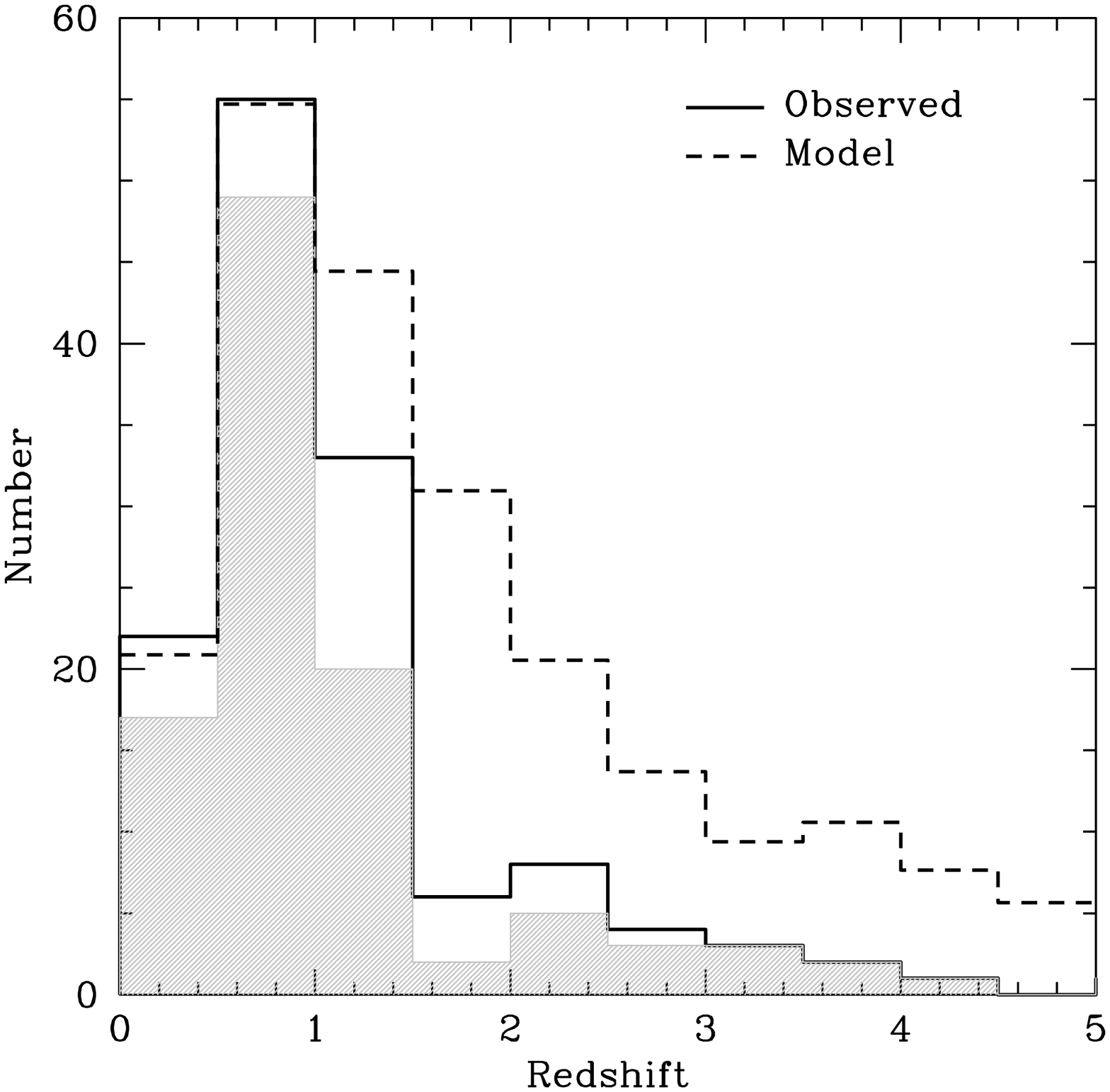}
\includegraphics[width=.45\textwidth]{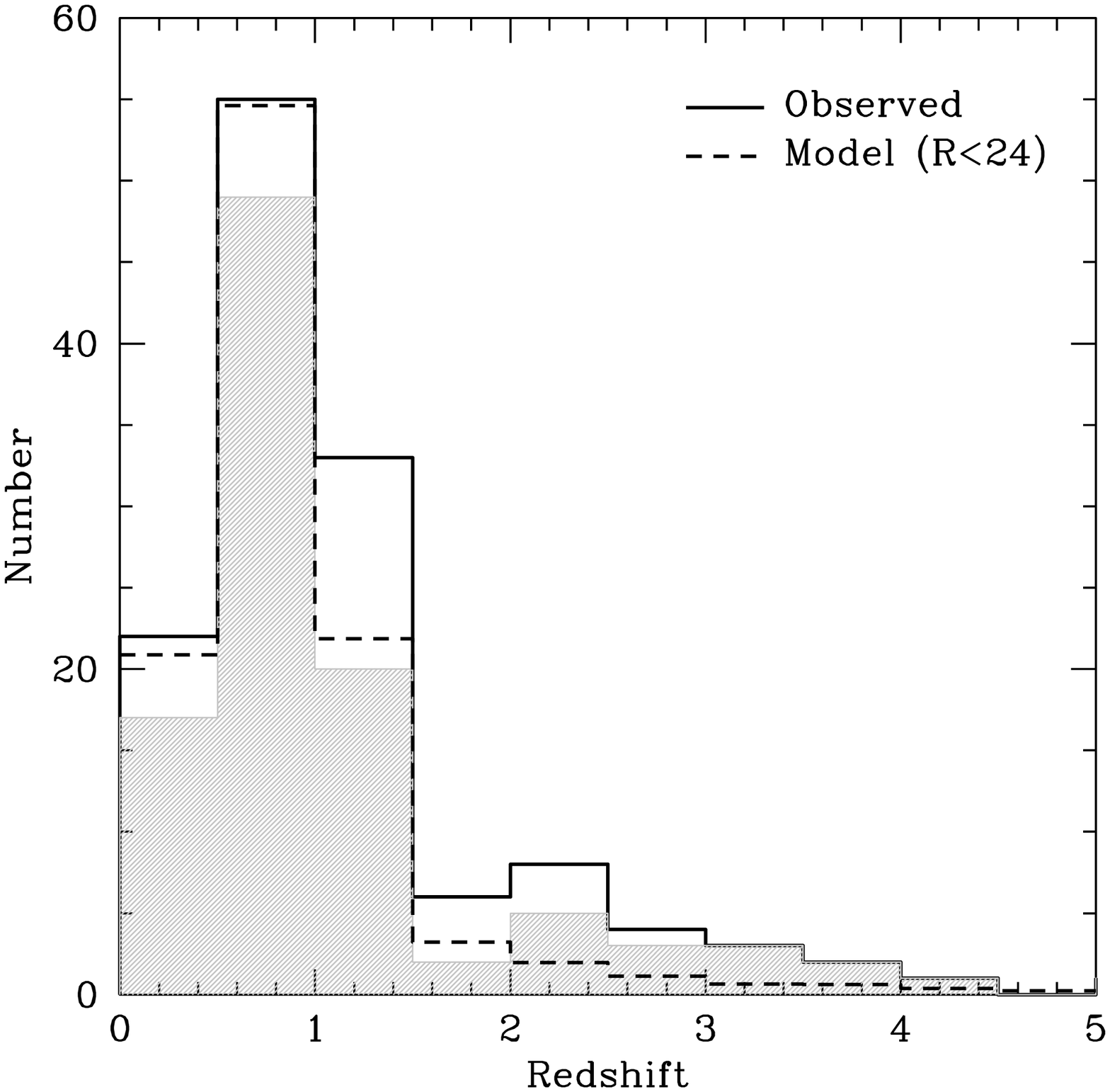}
  \caption{
({\it Left}) Observed spectroscopic redshift distribution 
({\it hatched histogram}) and photometric redshift 
distribution ({\it open histogram}) for the GOODS-North field,
compared to that expected from the unification model
({\it dashed line}). There is poor agreement at high redshifts.
({\it Right})
Imposing an optical flux limit on the model leads to
an expected distribution ({\it dashed line}) that agrees
very well with the observed distribution.}
  \label{figzdist}
\end{center}
\end{figure}

The unification-inspired model described above, 
like other population synthesis models (e.g., \citealp{gilli01}), 
predicts a large population of obscured AGN out to high redshift. 
The predicted peak of the redshift distribution 
({\it dashed histogram}, Fig.~\ref{figzdist}a) is near $z\sim1$,
not much higher than that observed in the GOODS data
({\it shaded/open histogram}, Fig.~\ref{figzdist}a), 
but there discrepancy at $z>1$.
Many of these high-redshift obscured AGN, because they are optically faint,  
are not identified in follow-up optical spectra of X-ray sources. 
This can be clearly seen imposing the spectroscopic limit, $R\sim24$~mag, 
on the expected redshift distribution 
({\it dashed histogram}, Fig.~\ref{figzdist}b);
in this case, the observed and expected distributions agree well
because the higher-redshift AGN remain largely unidentified. 
Equivalently, AGN without spectroscopic identifications 
will preferentially be faint, obscured sources.


The advent of multiple large AGN samples with high spectroscopic
completeness allowed \citet{barger05} to deduce a dependence of
obscured fraction on optical luminosity.
We incorporated a simple functional form of that dependence 
in our model --- namely, a linear transition between
100\% of AGN with $L_X = 10^{42}$~ergs/s being obscured
and no AGN with $L_X = 10^{46}$~ergs/s being obscured.
Taking into account the selection effects due to flux limits,
the model then matches the observed dependence very well 
(Fig.~\ref{figLumRatio}).
The resulting number counts and redshift distributions for GOODS
do not change, since only a restricted luminosity range 
(primarily $10^{43-44}$~ergs/s) is probed by that survey.
In all work subsequent to \citet{treister04} --- 
on infrared counts, X-ray background, and
evolution of obscuration (see below) --- 
we incorporated this luminosity dependence.

\begin{figure}
\begin{center}
\includegraphics[width=.30\textwidth]{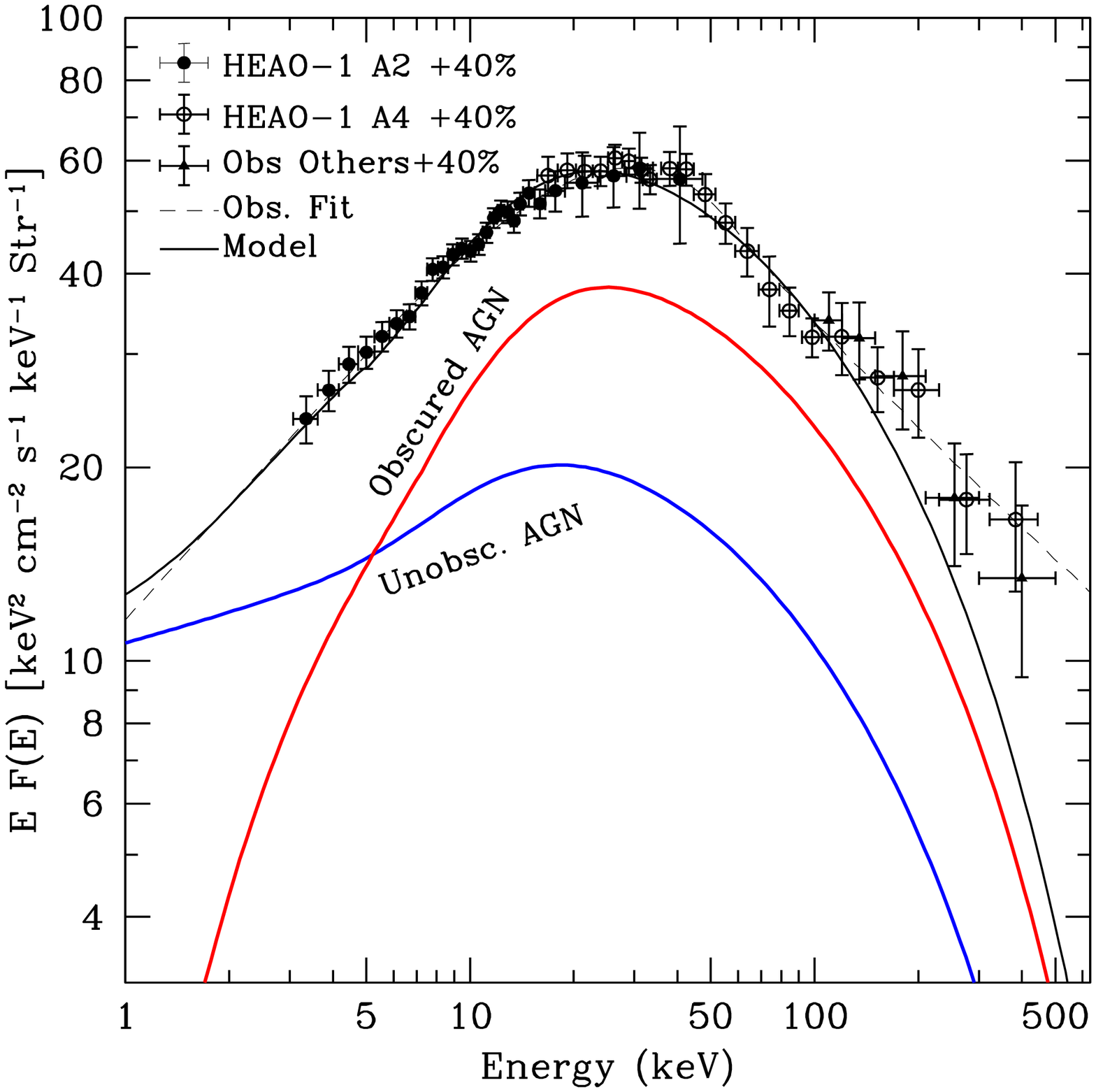}
\includegraphics[width=.30\textwidth]{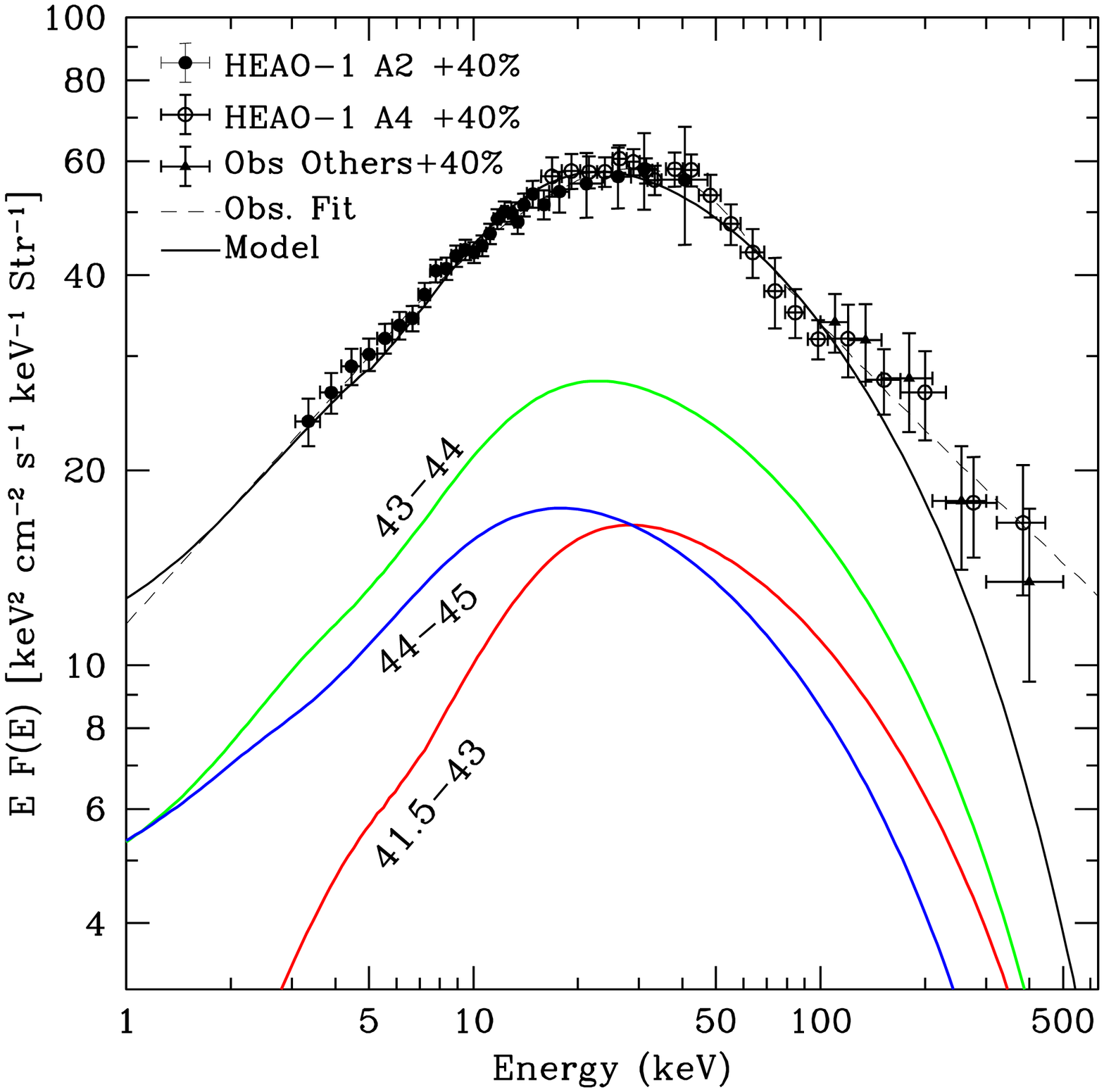}
\includegraphics[width=.30\textwidth]{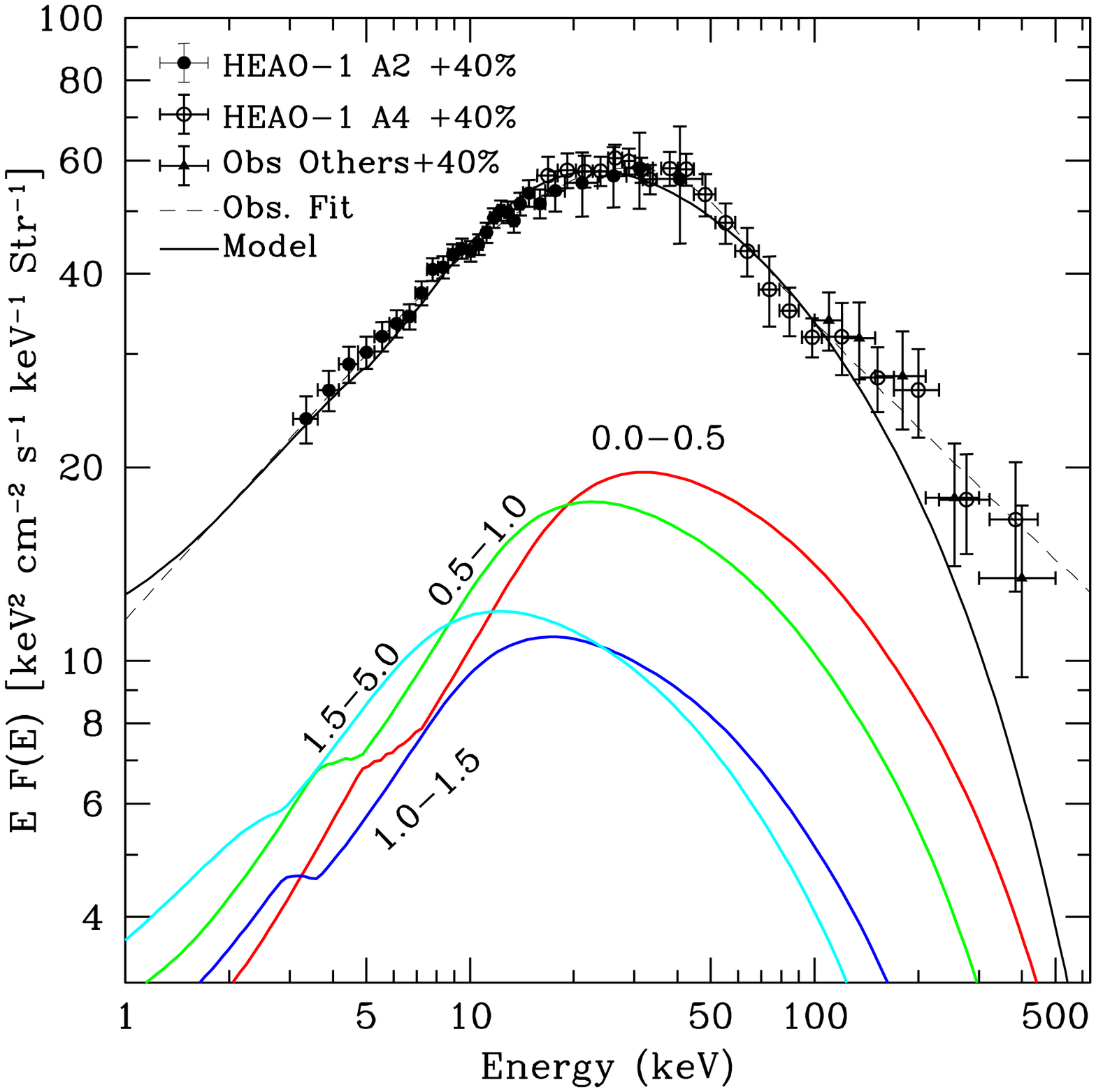}
  \caption{
The predicted X-ray background spectrum ({\it heavy line}; 
\citealp{treister05b})
from the unification-inspired population synthesis model
agrees very well with the observed data ({\it data points}).
({\it Left}) Obscured AGN ({\it thin, red, peaked curve}) dominate at high 
energies ($E>10$~keV), while unobscured AGN ({\it thin, blue, broad curve})
dominate in the Chandra and XMM bands 
($2~{\rm keV} < E < 10$~keV).
There is little freedom in the fit parameters, especially at 
low energies, so this agreement is remarkably good.
(Emission at $E>100$~keV is dominated by blazars, which are
not modeled here.)
The X-ray background is dominated by moderate luminosity
AGN, in the range $10^{43-44}$~ergs/s ({\it middle panel}) 
and by intermediate redshifts, $z\sim 0.5-1.5$
({\it right panel}).
}
  \label{figXRBGSynth}
\end{center}
\end{figure}

Our population synthesis model predicts, with very little freedom,
the spectrum of the X-ray background. 
We simply sum the X-ray emission from all the AGN.
There is some freedom in that we (like everyone else who does this
kind of calculation) have to model the X-ray spectrum.
We used an absorbed power law plus an
iron line and a Compton-reflection hump, 
consistent with the AGN spectra that are well measured locally; 
we also assumed solar abundances and used
the $N_H$ distribution described above for
$\log N_H = 20$-24~cm$^{-2}$. 
The $N_H$ distribution for higher column densities 
($N_H > 10^{24}$~cm$^{-2}$) is poorly constrained at present,
so we extrapolated from $N_H < 10^{24}$~cm$^{-2}$ 
with a flat slope (again, much like what others have assumed).
Beyond this, one can adjust the metallicity (this increases the hard X-rays,
holding everything else constant) or the include a small range
of power-law slopes (ditto; \citealp{gilli07}), 
but we did not feel the data could constrain those parameters and
so left them fixed.
In any case, the integral constraint of the X-ray background is an excellent
check on whether the demographics of AGN assumed here is realistic.
Figure~\ref{figXRBGSynth} shows that, with almost no free parameters, 
the data are very well fit indeed. 
(The normalization of the background at $E>10$~keV is discussed
below in \S~3.3.)

\begin{figure}
\begin{center}
\includegraphics[scale=0.4]{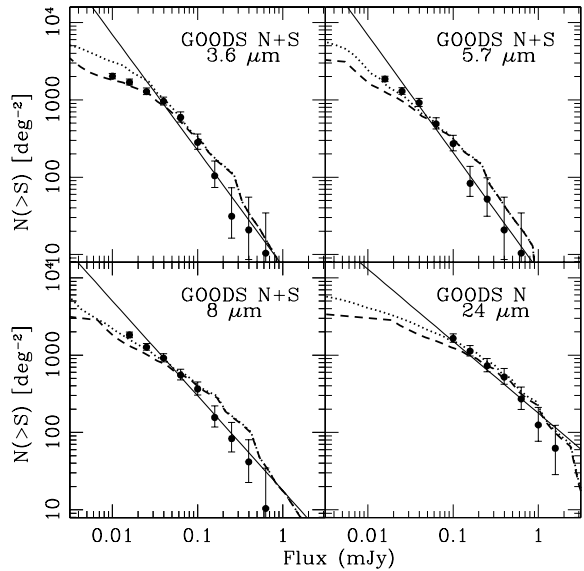}
  \caption{ 
({\it Filled circles}) Observed cumulative infrared flux distributions 
for X-ray detected AGN in the GOODS-North and South fields 
in three IRAC bands, and for the North field only in the MIPS 24-micron band. 
Error bars correspond to the 84\% confidence level in the 
number of sources in that bin \citep{gehrels86}. 
Because this is a cumulative plot, the bins are not independent. 
({\it Solid line}) Power-law fit to the observed number counts. 
({\it Dashed line}) Expected infrared flux distribution 
\citep{treister06a} based on our population synthesis model,
taking into account the X-ray flux limit in the GOODS fields 
and the luminosity dependence of the obscured AGN fraction,
and correcting for the effects of undersampling at the bright
end caused by the small volume of the GOODS fields.
({\it Dotted line}) Expected infrared flux distribution 
not considering the X-ray flux limit (i.e., all sources 
expected from extrapolating the AGN luminosity function). 
In general, the model explains the normalization and shape 
of the counts quite well, especially considering the 
poor statistics at the bright end.
At the faint end, where the host galaxy light dominates, 
the model diverges more significantly, 
since the model assumptions (no distribution 
in host galaxy magnitude or spectrum) are clearly far too simple.
}
 \label{figIRCounts}
\end{center}
\end{figure}

Another strong prediction of the Treister model 
concerns the infrared data, which at the time the model
was developed had not yet been taken.
We calculated the expected Spitzer counts in IRAC and 
MIPS 24-micron bands; Figure~\ref{figIRCounts} shows the 
excellent agreement with observations \citep{treister06a}. 
Small discrepancies at the faint end are due to the 
overly simple assumption of a single host galaxy magnitude
(which dominates at low fluxes), rather than a distribution.

\begin{figure}
\begin{center}
\includegraphics[width=.47\textwidth]{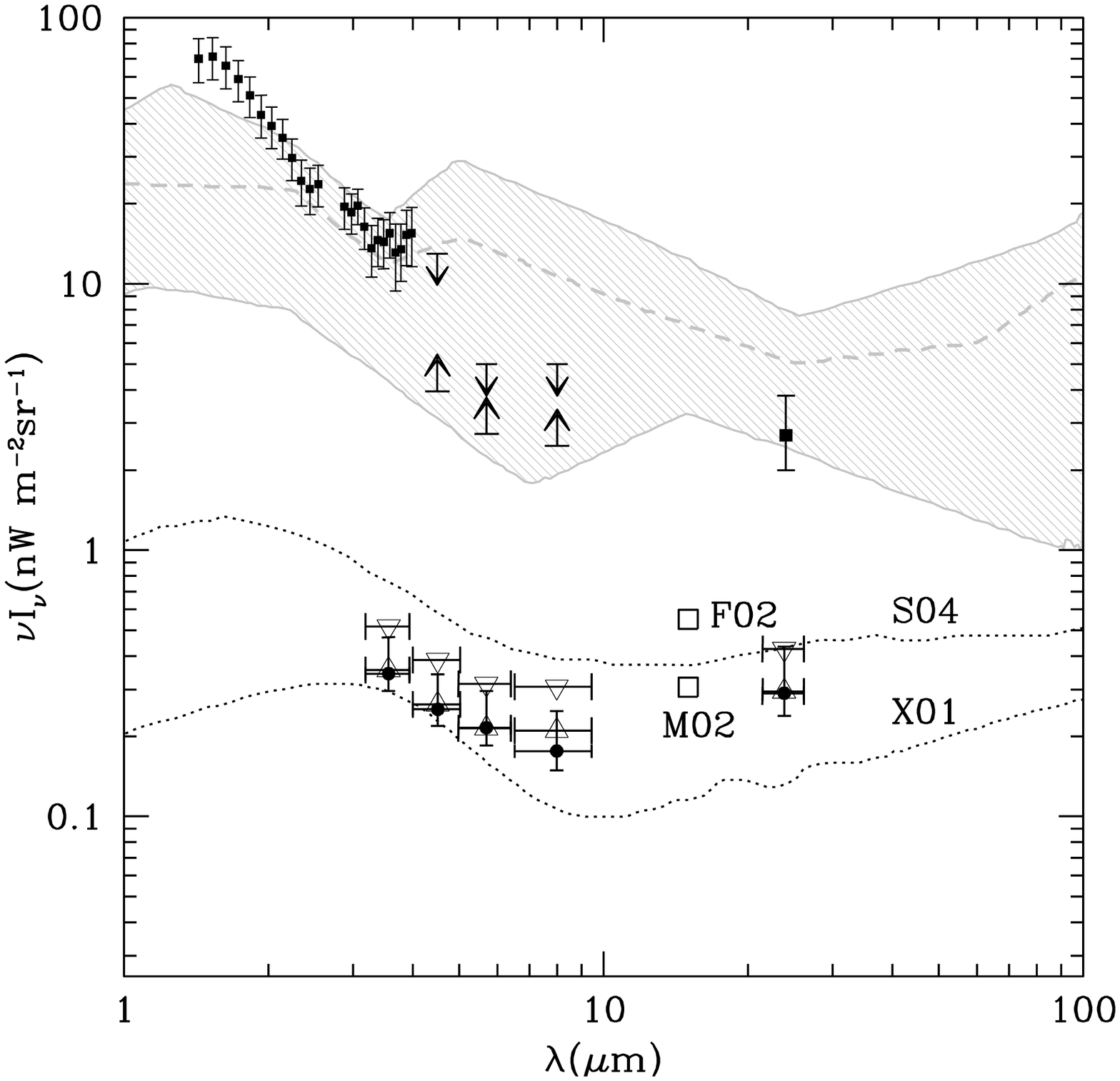}
\includegraphics[width=.47\textwidth]{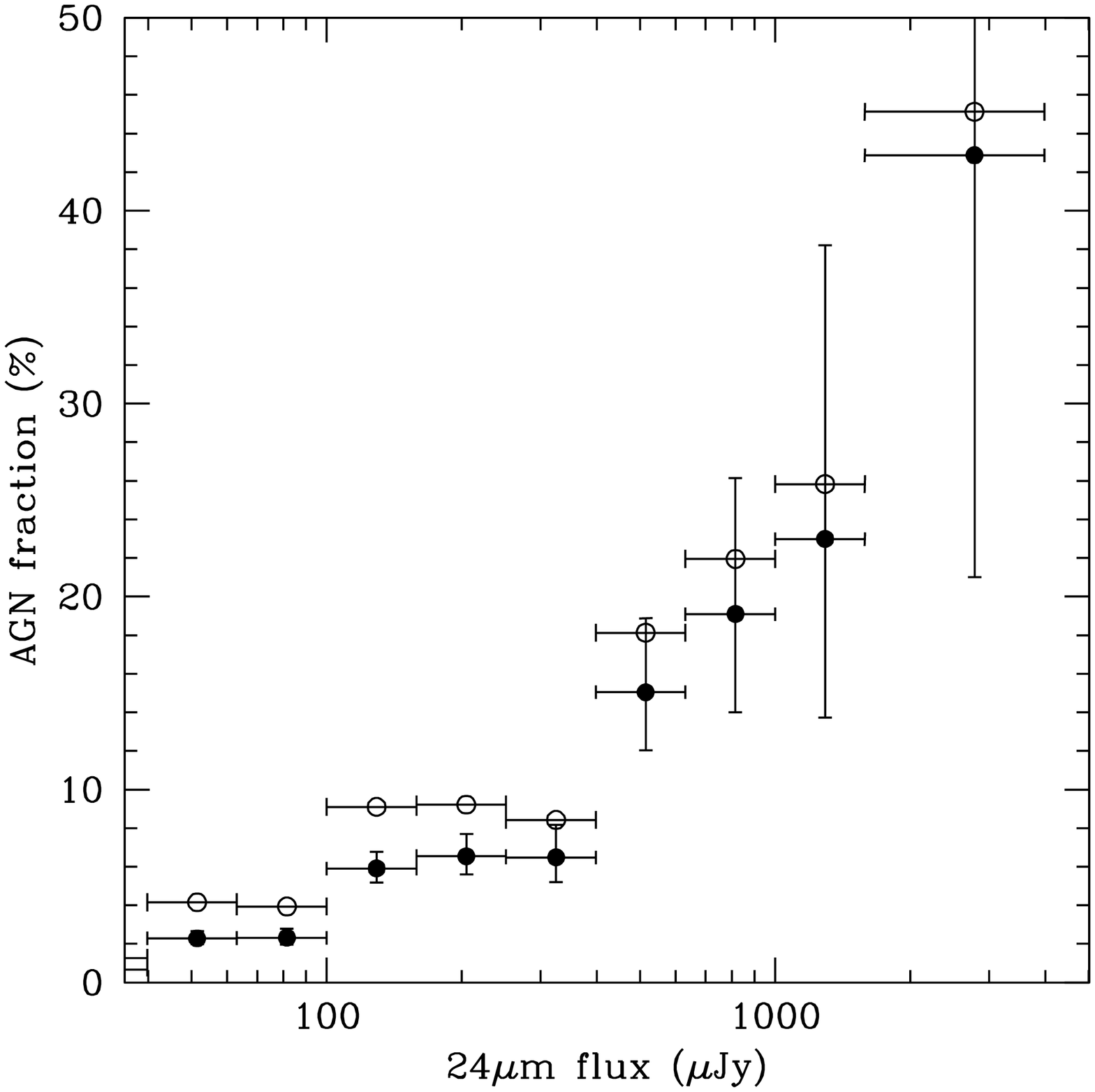}
  \caption{
({\it Left}) 
Extragalactic infrared background intensity as a function of wavelength. 
({\it Shaded region}) Allowed values for the total intensity
compiled by \citet{hauser01}. 
1-4~$\mu$m: Measurements reported by \citet{matsumoto05}.
4-8~$\mu$m: Lower limits from \citet{fazio04b}, 
upper limits from \citet{kashlinsky96} at 4.5~$\mu$m 
and \citet{stanev98} at 5.8 and 8~$\mu$m.
Measurement at 24~$\mu$m from \citet{papovich04}. 
({\it Solid circles}) Integrated infrared emission from
X-ray-detected AGN in the GOODS fields,
with 1$\sigma$ error bars from source number statistics
(which dominate over measured flux errors)
calculated as in \citet{gehrels86}.
({\it Triangles}) Integrated AGN emission from the models of \citet{treister05b} if only X-ray-detected AGN are considered
({\it lower}) or including all AGN ({\it upper}).
({\it Dotted lines}) 
Expected AGN integrated emission from the AGN models
of Silva {\it et al.} (2004; {\it upper line}) and 
Xu {\it et al.} (2001; {\it lower line}).
({\it Open squares})
Expected AGN integrated emission at 15~$\mu$m 
from the extrapolation to fainter fluxes of IRAS and ISO 
(\citealp{matute02} [M02]; corrected by a factor of 5 
to include the contribution from obscured AGN) 
and ISO only (\citealp{fadda02} [F02]) observations.
({\it Right}) 
The fraction of sources that are AGN rises sharply
with 24~$\mu$m infrared flux ({\it filled circles}).
Vertical error bars show the 1$\sigma$ 
Poissonian errors on the number of sources \citep{gehrels86}.
Also shown is the contribution corrected 
by the AGN expected to be missed by X-ray selection
({\it open circles}), 
as estimated using our AGN population synthesis model;
this shows the same strong dependence on infrared flux, 
indicating that that dependence is not a selection effect 
induced by the X-ray flux limit.
}
  \label{figIRBG}
\end{center}
\end{figure}

From the Spitzer observations, we calculate the minimum AGN contribution 
to the extragalactic infrared background, 
obtaining a lower value than previously estimated, 
ranging from 2\% to 10\% in the 3-24 micron range \citep{treister06a}. 
Accounting for heavily obscured AGN that, 
according to our population synthesis model, are not detected in X-rays, 
the AGN contribution to the infrared background increases by $\sim$45\%, 
to $\sim$3-15\%. 
Figure~\ref{figIRBG}a shows the AGN contributions deduced from 
GOODS data \citep{treister06a} compared to other estimates and 
to the (uncertain) total extragalactic background light, 
as a function of infrared wavelength.
The GOODS measurements place the strongest constraints to date on the
AGN contribution to the extragalactic background light, 
indicating that stars dominate completely over AGN at infrared wavelengths.

The fraction of sources that are AGN rises sharply
with 24~$\mu$m infrared flux (Fig.~\ref{figIRBG}b).
Thus in deep surveys like GOODS or COSMOS, 
AGN are a small fraction of the total infrared source population,
while in high flux limit surveys like SWIRE \citep{lonsdale03},
they constitute a much higher fraction. 
AGN detected in large-area, shallow surveys are on average 
closer and/or more luminous than AGN found in deep pencil-beam surveys.
We show in Figure~\ref{figIRBG}b that this 
very strong dependence of AGN fraction on infrared flux 
is not an artifact of the X-ray flux limit.
Specifically, using our AGN population synthesis model, 
we plot the number of AGN as a function of infrared flux
including those too faint to be detected in X-rays 
in the Chandra Deep Fields ({\it open circles}). 
Clearly the trend is independent of X-ray flux limit.

\subsection{Integral and Swift Hard X-Ray Surveys}

If the population synthesis model presented here is correct, 
$\sim$50\% of AGN are currently missed even in deep X-ray
surveys with {\it Chandra} or {\it XMM}. 
These are very obscured AGN, many of them Compton thick 
(i.e., $N_H > 10^{24}$~cm$^{-2}$), 
so all but the hardest X-ray surveys are biased against them.
Fortunately, hard X-ray instruments on the INTEGRAL and Swift satellites
can now reach fluxes below $\sim$10$^{-11}$~ergs~cm$^{-2}$~s$^{-1}$ 
at energies above 20~keV. 
Serendipitous AGN surveys at these energies covering almost 
the full sky have been carried out using Swift/BAT \citep{markwardt05} 
and INTEGRAL/IBIS \citep{beckmann06,sazonov07}, 
yielding samples of $\sim$100 AGN each. 
These surveys provide an unbiased view of the AGN population, 
independent of the amount of obscuration, although at present
the sensitivity is sufficient to reach only local populations.

\begin{figure}
\begin{center}
\includegraphics[width=.45\textwidth]{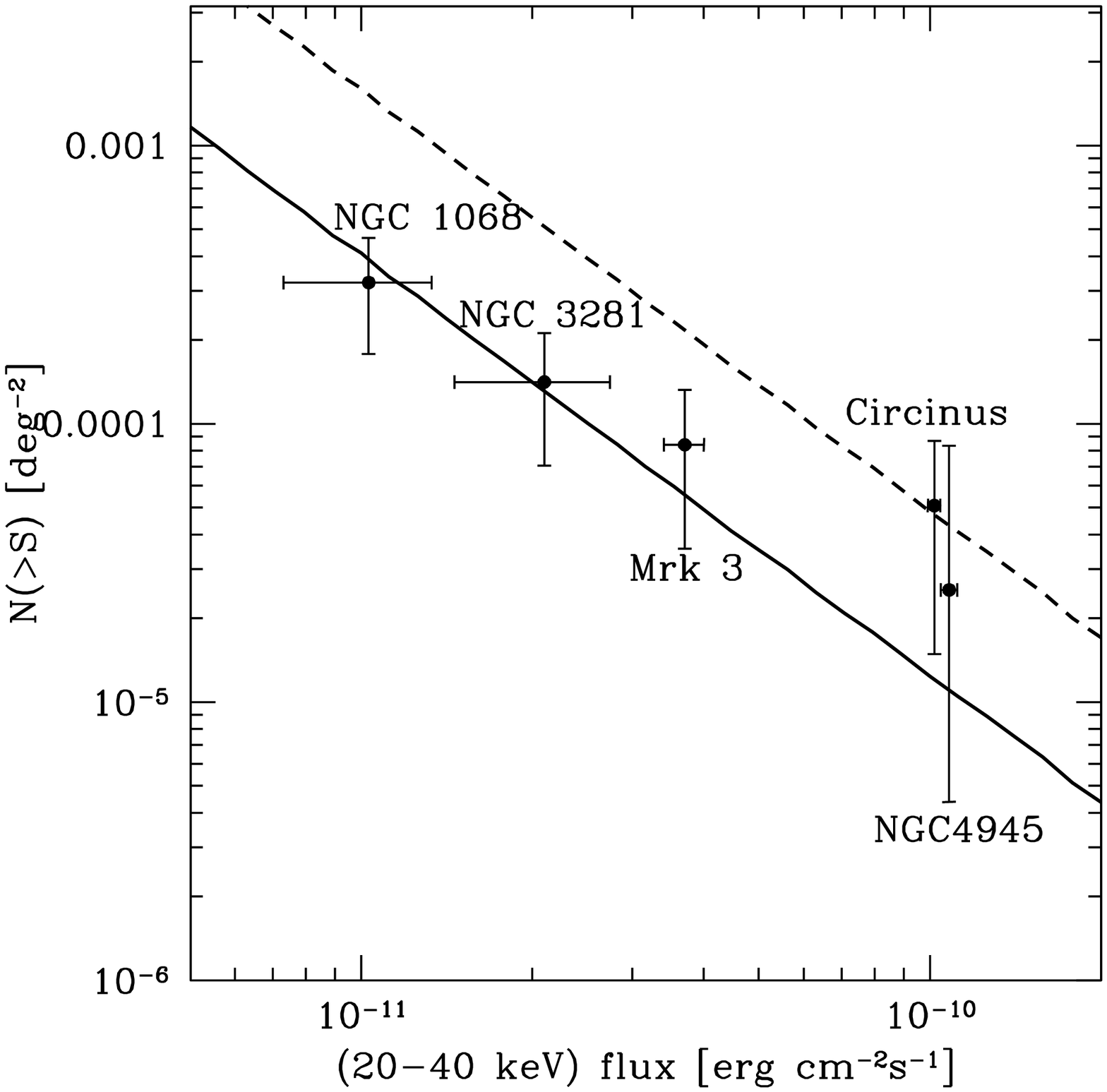}
\includegraphics[width=.45\textwidth]{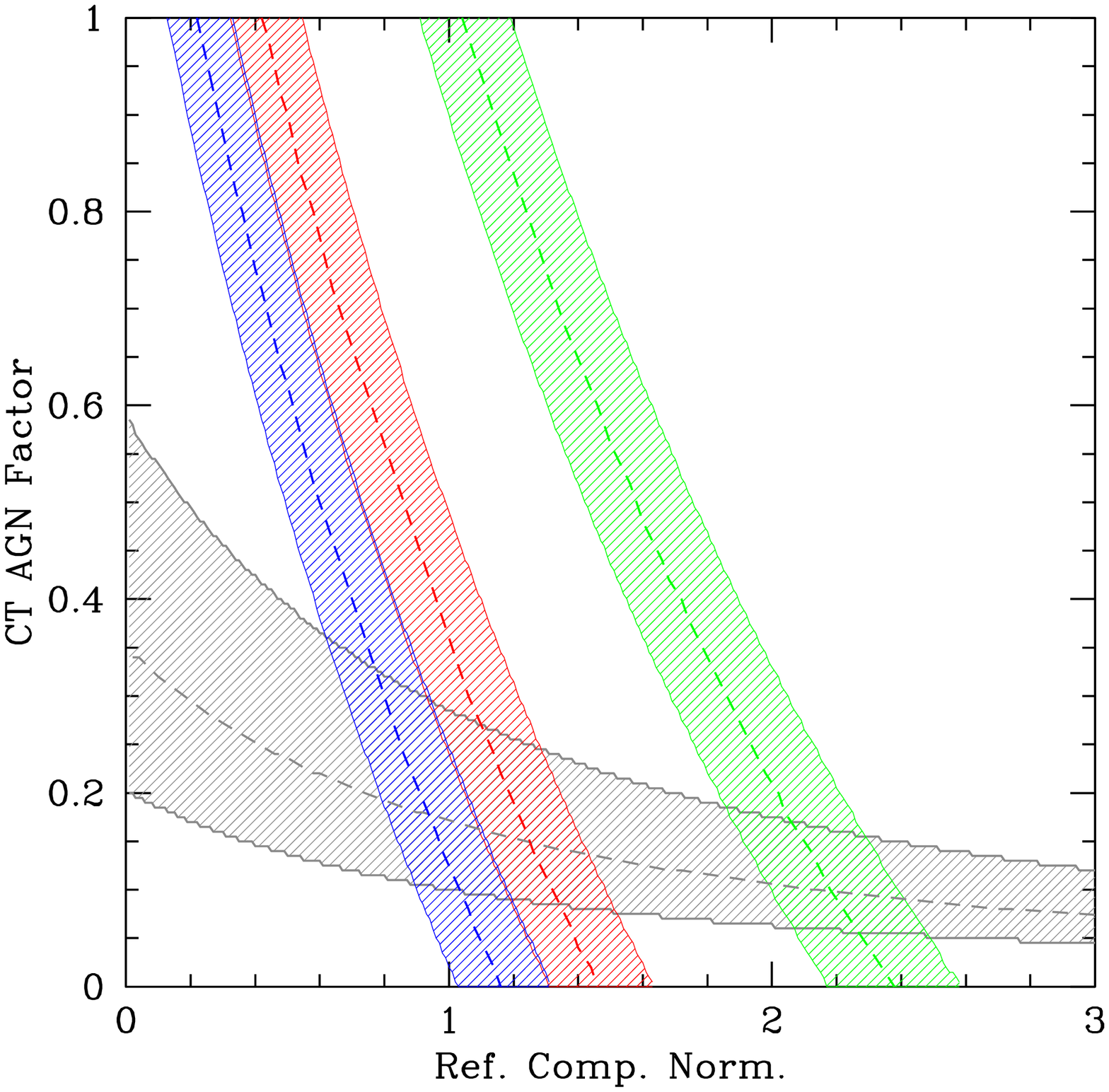}
\caption{({\it Left}) LogN-$\log$S relation for Compton-thick AGN only. 
The {\it dashed line} shows the original prediction of \citet{treister05b}, 
which assumed a reflection component normalization of 1 
and a Compton-thick AGN factor of 1 (i.e., as many Compton-thick AGN
as Compton-thin AGN in the next lowest decade of the $N_H$ distribution). 
Data points are from the INTEGRAL survey of \citet{sazonov07}. 
The population synthesis model can be brought to agreement with 
the observations if there are a factor of 4 fewer Compton-thick AGN 
(i.e., the Compton-thick AGN factor is 0.25).
({\it Right}) Compton-thick AGN factor versus normalization 
of the Compton reflection component. 
The {\it gray region} shows the values of these parameters 
that produce a space density of Compton-thick AGN consistent 
with the observed value in the \citet{sazonov07} sample, 
considering 1-$\sigma$ statistical fluctuations. 
The allowed values of the parameters needed to fit 
the intensity of the X-ray background at 20-40~keV, 
assuming a 5\% uncertainty in each case, are shown by
the {\it blue region} for the original \citet{gruber99} measurements, 
the {\it red region} for the INTEGRAL values and 
by the {\it green region} for the \citet{gruber99} points increased by 40\%
(which was the assumption in \citealp{treister05b}). 
A large value of the reflection component is required 
to obtain values consistent with the latter X-ray background intensity, 
inconsistent with observations of the reflection component in individual AGN.
Both the reflection value and the renormalization of the 
\citet{gruber99} X-ray background intensity must be lower.
}
\label{figCT}
\end{center}
\end{figure}

Figure~\ref{figCT}a shows the $\log$N-$\log$S distribution 
from the INTEGRAL catalog \citep{sazonov07} of AGN 
confirmed as Compton thick with $N_H$ measurements from X-ray data. 
Clearly the original population synthesis model 
of \citet{treister05b} overpredicted the density of 
Compton-thick AGN by a factor of $\sim$4. 
This is not too surprising, as the model included assumptions that,
at the time of the publication of that work, were unconstrained ---
for example, the reflection fraction\footnote{The reflection fraction
is the geometrical factor describing the solid angle,
relative to $2\pi$, subtended by cold reflecting material 
as seen from the primary X-ray source.} was assumed to be unity,
and the number of Compton-thick AGN was simply extrapolated 
with a flat slope from the $N_H$ distribution at lower column densities.

Now we explore the constraints on these quantities that come
from the new Swift and INTEGRAL surveys.
First, we define a ``Compton-thick AGN correction factor,"
which simply multiplies the original flat-extrapolation assumption
to match the observed number density of AGN with $N_H>10^{24}$~cm$^{-2}$. 
According to a $\chi^2$ minimization, 
the best-fit value for the Compton-thick AGN factor is 0.25;
this produces a good agreement between model and observations, 
with a reduced $\chi^2$ of 0.3.

Second, we consider the ratio of direct and reflected X-rays.
The spectrum of Compton-thick AGN at high energies 
is dominated by the Compton reflection component (e.g., \citealp{matt00}), 
which has a strong peak at $E\sim$30~keV \citep{magdziarz95}. 
The observed spectrum of the X-ray background, 
which we now understand as the integrated emission from previously unresolved AGN, 
also has peak at about the same energy \citep{gruber99}. 
The normalization of the reflection component relative 
to the direct emission is known only for a few nearby AGN, 
mostly from BeppoSAX observations (e.g., \citealp{perola02}), 
and therefore this parameter is usually taken as an assumption 
in AGN population synthesis models that can explain 
the spectral shape and intensity of the X-ray background. 
The resulting peak X-ray background intensity depends 
on both the assumed space density of Compton-thick AGN 
and the normalization of the reflection component, $R$;
while satisfying the overall intensity constraint, one can trade
increased reflection for fewer Compton-thick AGN, or vice versa.
Figure~\ref{figCT}b shows the allowed regions 
in terms of the Compton-thick AGN factor versus reflection parameter. 
That is, the shaded regions show the values of these two parameters 
that produce an integrated X-ray background intensity in the 20-40~keV range, 
the region most affected by both the number of Compton-thick AGN and $R$, 
consistent with:
({\it green region})
the \citet{gruber99} data increased by 40\%
(the maximum suggested renormalization to match higher
estimates with imaging instruments at lower energies;
\citealp{treister05b}); 
({\it red region})
the recent INTEGRAL X-ray background intensity measured 
using Earth occultation by \citet{churazov06};
({\it blue region})
the original \citet{gruber99} measurements, not renormalized. 
In each case uncertainties in the X-ray background intensity
were assumed to be 5\%. 
The gray shaded region shows the parameter space that 
produces a density of Compton-thick AGN consistent 
with the observations in Fig.~\ref{figCT}a.

Previously, \citet{treister05b} assumed an 
X-ray background intensity consistent with the
\citet{gruber99} value increased by 40\%, 
which was well fit by a Compton-thick AGN factor of 1
(flat extrapolation of the $N_H$ distribution; Fig.~\ref{figXRBGSynth}). 
However, such a high value of the Compton-thick AGN factor 
is clearly inconsistent with the new observational constraints
on the density of Compton-thick AGN.
So, either the intensity of the X-ray background is lower, 
as suggested by recent INTEGRAL measurements \citep{churazov06}, 
or the average value of the reflection parameter is high, $R\sim$2,
or some combination of the two. 
Observations of individual sources seem to indicate that 
such a high value for the reflection component is unlikely. From 
a sample of 22 Seyfert galaxies, excluding Compton-thick sources, 
\citet{malizia03} concluded that both obscured and unobscured sources 
have similar reflections component with normalization values in the 0.6-1 range. 
A similar value of $R\simeq$1 was reported by \citet{perola02} 
based on BeppoSAX observations of a sample of nine Seyfert 1 galaxies. 
Although with large scatter, 
normalizations for the average reflection component of 
0.9 for Seyfert 1 and 1.5 for Seyfert 2 
were measured from BeppoSAX observations 
of a sample of 36 sources \citep{deluit03}. 
Therefore, a value of $R\sim$1 for 
the normalization of the reflection component, 
required by both the observed Compton-thick AGN space density 
and the X-ray background intensity reported by INTEGRAL and HEAO-1,
is in good agreement with the observed values in nearby Seyfert galaxies. 

\begin{figure}
\begin{center}
\includegraphics[width=.45\textwidth]{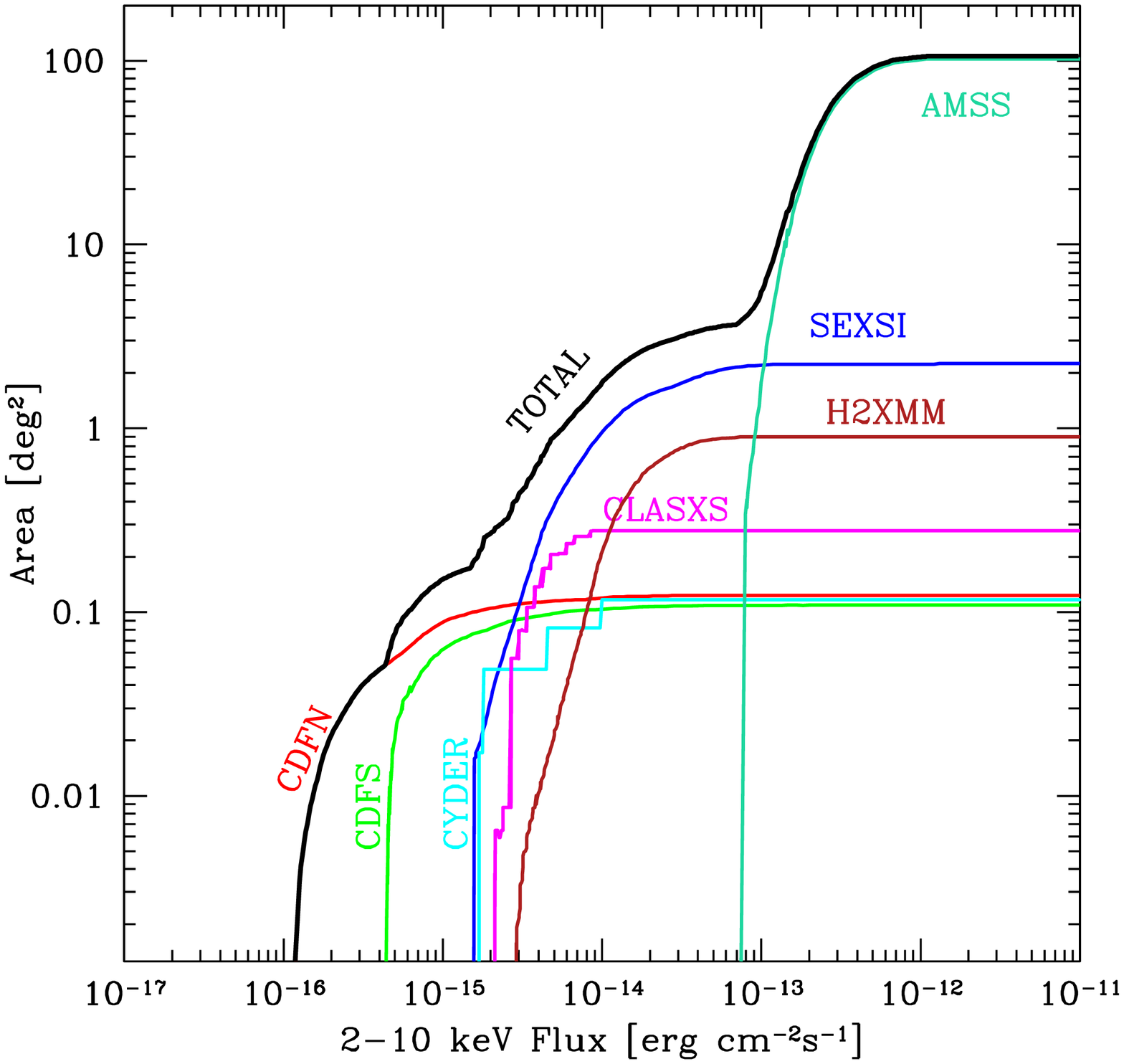}
\includegraphics[width=.45\textwidth]{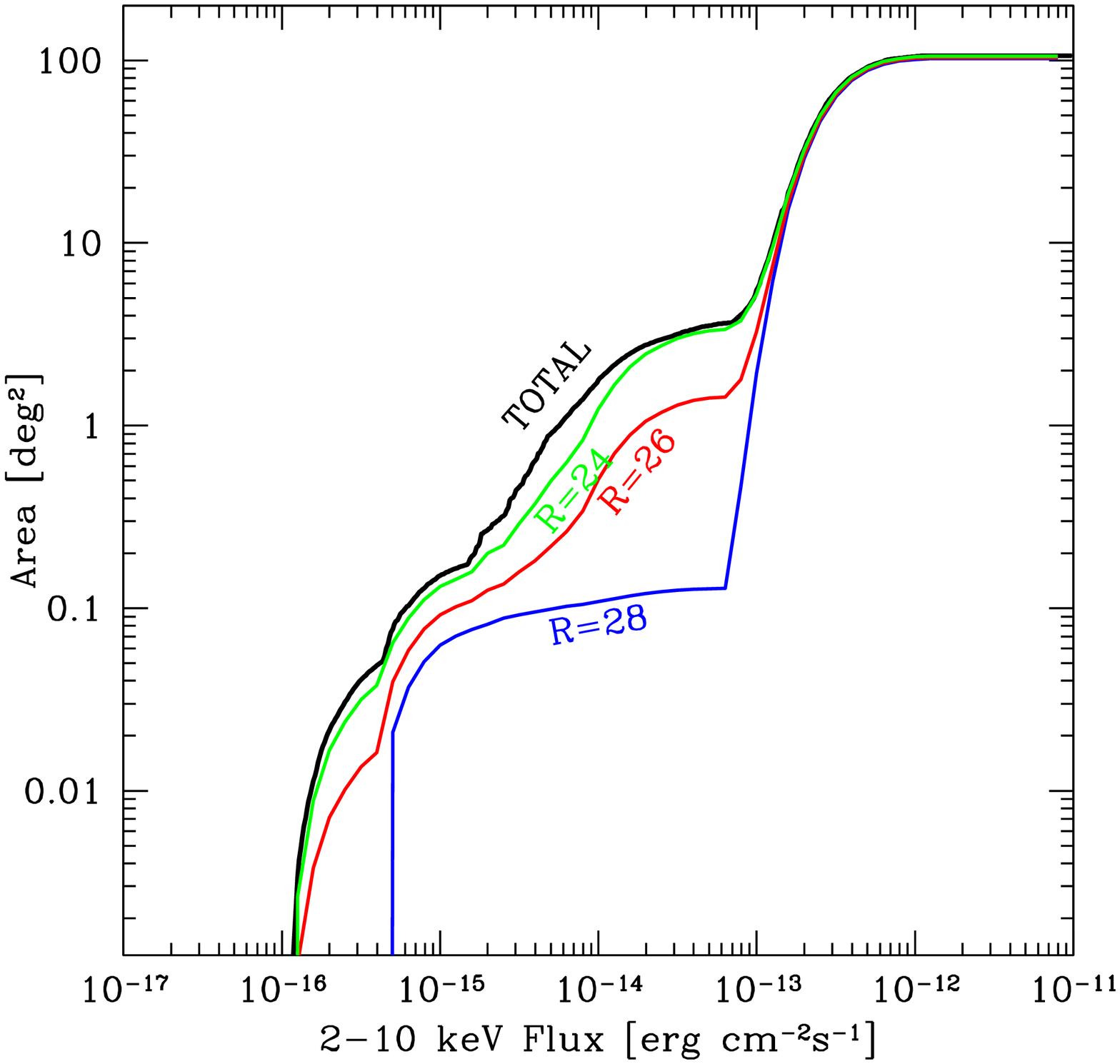}
\caption{({\it Left}) Area versus hard X-ray flux for 
each of the seven surveys combined in this work ({\it light colored lines}) 
and for the total sample ({\it thick black line}). 
({\it Right}) Total effective area as a function of 
X-ray flux and optical magnitude (for $R$=24, 26 and 28~mag), 
taking into account the spectroscopic incompleteness 
of each survey (see text for details). 
These curves are used to compute the expected fraction 
of identified obscured AGN for an intrinsically non-evolving ratio.
}
\label{figREvolArea}
\end{center}
\end{figure}

\section{Evolution of the Obscured Fraction of AGN} 

The X-ray background, being an integral constraint, 
is not a strong probe of the fraction of AGN that is obscured
(as we showed in the previous section),
much less of the evolution of that fraction with cosmic epoch. 
As shown in Figure~\ref{figXRBGSynth}, 
a simple population synthesis model 
in which obscured and unobscured AGN have the same evolution 
(or equivalently, the fraction of AGN that is obscured does not evolve)
is fully consistent with the spectrum of the X-ray background. 
Now, however, the large X-ray samples that have become 
available in the past few years, 
spanning a range of depths and with high spectroscopic completeness,
allow us to determine whether and how the fraction of 
obscured AGN depends on redshift \citep{treister06b}.

To study the evolution of the obscured AGN fraction, one needs 
to distinguish between the effects of redshift and luminosity,
which are correlated in any flux-limited sample, 
Wide area, shallow X-ray surveys (e.g., XBOOTES; \citealp{hickox06a})
sample moderate luminosity AGN at low redshifts and 
only high luminosity sources up to high redshifts, 
while deep pencil-beam surveys (e.g., CDFS; \citealp{giacconi02})
find moderate luminosity AGN out to high redshifts 
but lack rare, high-luminosity sources because of the small volume sampled. 
Combining the two extremes covers the luminosity-redshift plane effectively. 

\begin{figure}
\begin{center}
\includegraphics[width=\textwidth]{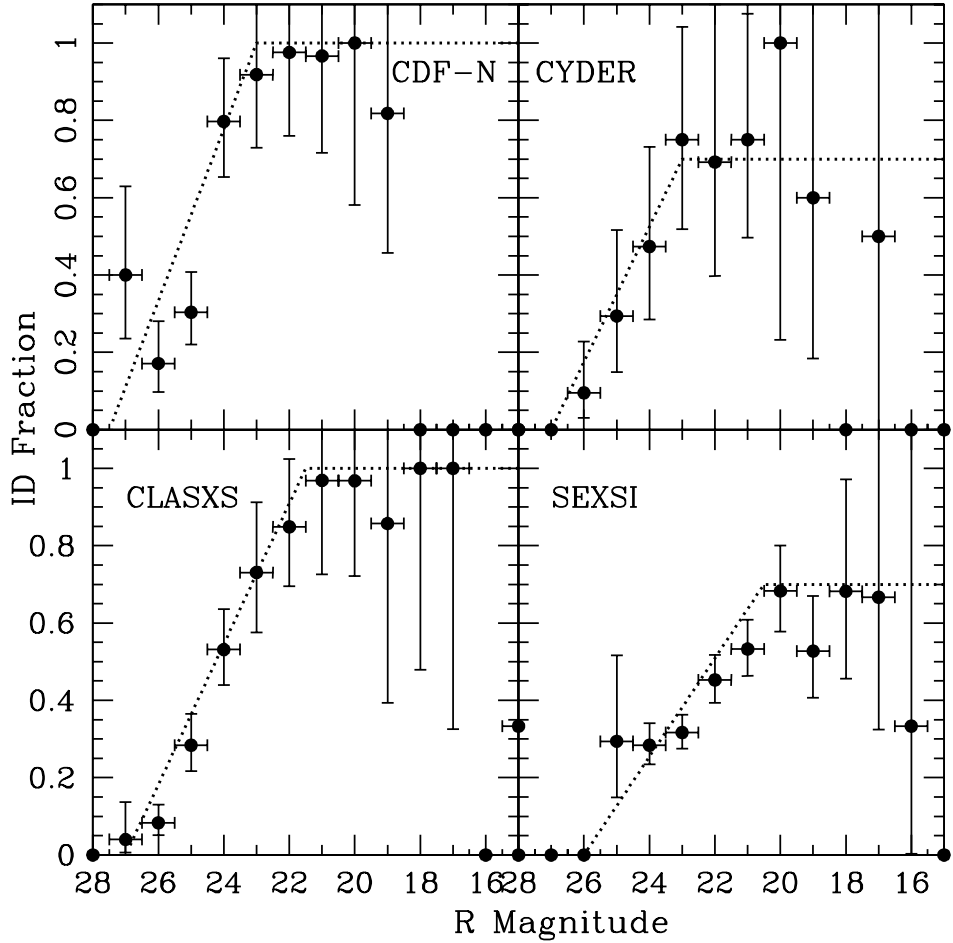}
\caption{Fraction of optically identified AGN as a function 
of optical magnitude ({\it data points}), 
for four of the seven surveys comprising our super-sample.
All seven are well described by a simple linear increase to
a constant fraction at bright magnitudes ({\it lines}).
This allows us to derive the effective survey area 
as a function of both X-ray flux and optical magnitude.
}
\label{figIDfrac}
\end{center}
\end{figure}

We generate an AGN super-sample comprising
seven large surveys with high identification fractions:
AMSS \citep{akiyama03},
SEXSI \citep{harrison03}, 
HELLAS2XMM \citep{baldi02}, 
CLASXS \citep{yang04}, 
CYDER \citep{treister05a}, 
and the two Chandra deep fields (\citealp{giacconi02,alexander03}).
This super-sample contains a total of 2341 hard X-ray-selected AGN
(X-ray sources with $L_X > 10^{42}$~ergs/s),
the largest such sample to date by a factor of $\sim$4 \citep{treister06b}.
It spans a range of luminosities at each redshift,
over a broad redshift range.
The total area of this super-sample as a function of X-ray flux 
is shown in Figure~\ref{figREvolArea}a. 

More than half the super-sample is optically identified.
We define an AGN as unobscured when there is evidence 
for broad emission lines in their spectra, 
and as obscured AGN otherwise.
The ``obscured fraction" is then the number of 
obscured AGN divided by the total number of AGN.
For obvious reasons, the identified fraction of any survey
depends on the brightness of the optical counterparts. 
We parameterized the identified fraction of each of the seven surveys
with a simple function that is constant at bright magnitudes and 
falls linearly to faint magnitudes 
(Fig.~\ref{figIDfrac}; see \citealp{treister06b} for details).
We then weight the area versus X-ray flux curve (Fig.~\ref{figREvolArea}a)
by the completeness at each optical magnitude, for each survey,
and sum those to get the effective area of the super-sample
as a function of both X-ray flux and optical magnitude
(Fig.~\ref{figREvolArea}b).
This allows us to calculate the expected numbers of 
optically identified AGN for the super-sample, 
i.e., we can now correct for both X-ray and optical spectroscopic limits.

\begin{figure}
\begin{center}
\includegraphics[scale=0.4]{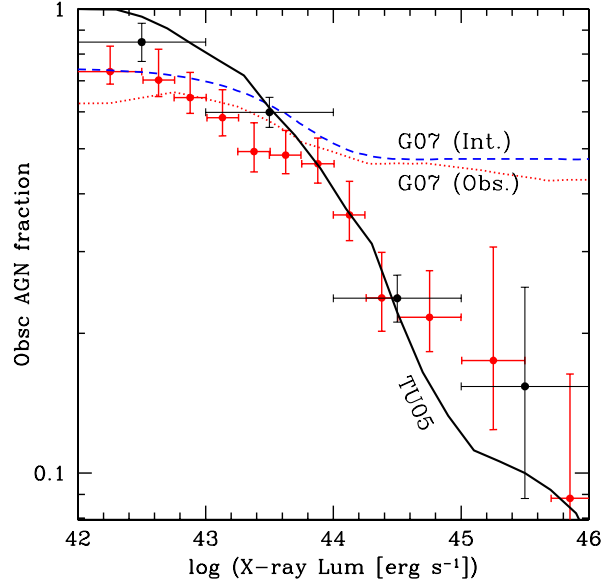}
\caption{Fraction of obscured AGN in the super-sample
as a function of hard X-ray luminosity. 
({\it Black data points}) Observed distribution for 
the super-sample described in this work. 
({\it Red data points}) Compilation of G. Hasinger (2007, in prep.) 
used in the work reported by \citet{gilli07}. 
({\it Black solid line})
Expected luminosity dependence for the Treister population synthesis models,
taking into account the X-ray flux limit and 
spectroscopic completeness of the super-sample
described here \citep{treister06b}. 
({\it Blue dashed and red dotted lines})
Intrinsic and bias-corrected (i.e., expected) luminosity dependences
for the population synthesis model of \citet{gilli07}.
This assumed luminosity dependence is not a good description 
of AGN demographics at higher luminosities;
the \citealp{gilli07} model is constrained primarily by
the X-ray background intensity, 
which is dominated by moderate luminosity AGN. 
}
\label{figLumRatio}
\end{center}
\end{figure}

As discussed earlier (\S~\ref{secResults}), 
the fraction of obscured AGN depends on luminosity \citep{barger05}.
Our population synthesis model assumed a linear transition between
100\% obscured fraction at $L_X = 10^{42}$~ergs/s 
and 0 obscured fraction at $L_X = 10^{46}$~ergs/s.
Taking into account the selection effects due to flux limits
(Fig.~\ref{figREvolArea}b),
this assumption ({\it black line}, Fig.~\ref{figLumRatio})
matches the observed dependence very well 
in the present super-sample
({\it black line}, Fig.~\ref{figLumRatio}).
The somewhat shallower luminosity dependence adopted by 
(\citealp{gilli07}; {\it blue dashed line} is their assumed dependence,
{\it red dotted line} incorporates flux limits),
in contrast, does not describe the obscured fraction well
for high-luminosity AGN, probably because their model was constrained 
by the X-ray background, which is dominated by moderate luminosity AGN.

Our population synthesis model fits 
the X-ray, optical, and infrared counts of AGN;
the X-ray background (modulo the trade-off between the number of
Compton-thick AGN and the reflection fraction of each);
the hard X-ray counts measured with Swift and INTEGRAL (ditto);
and the observed luminosity dependence of the obscured fraction of
AGN in the largest AGN sample to date.
What can it tell us about the evolution of this obscured fraction?

\begin{figure}
\begin{center}
\includegraphics[width=.45\textwidth]{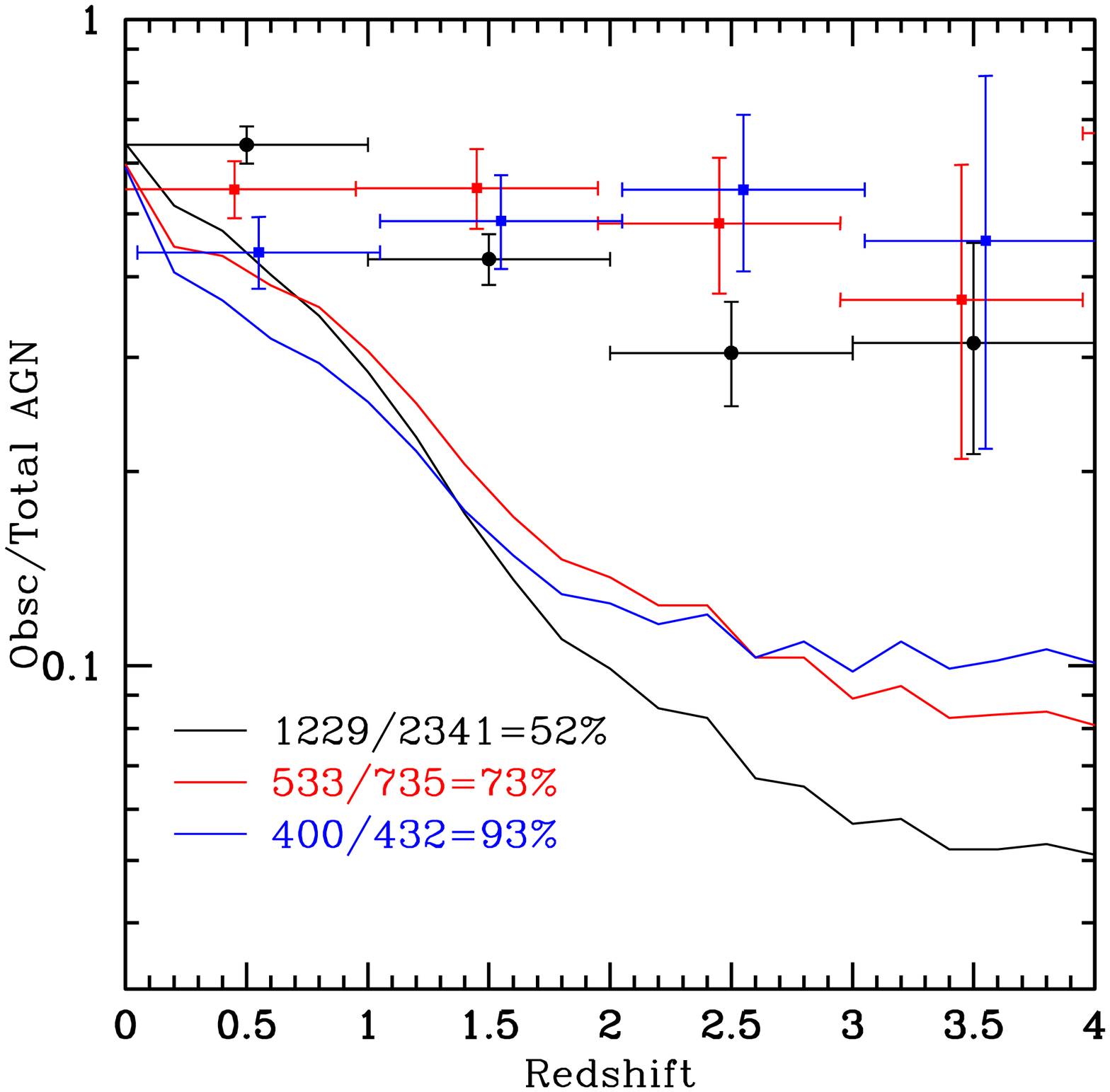}
\includegraphics[width=.45\textwidth]{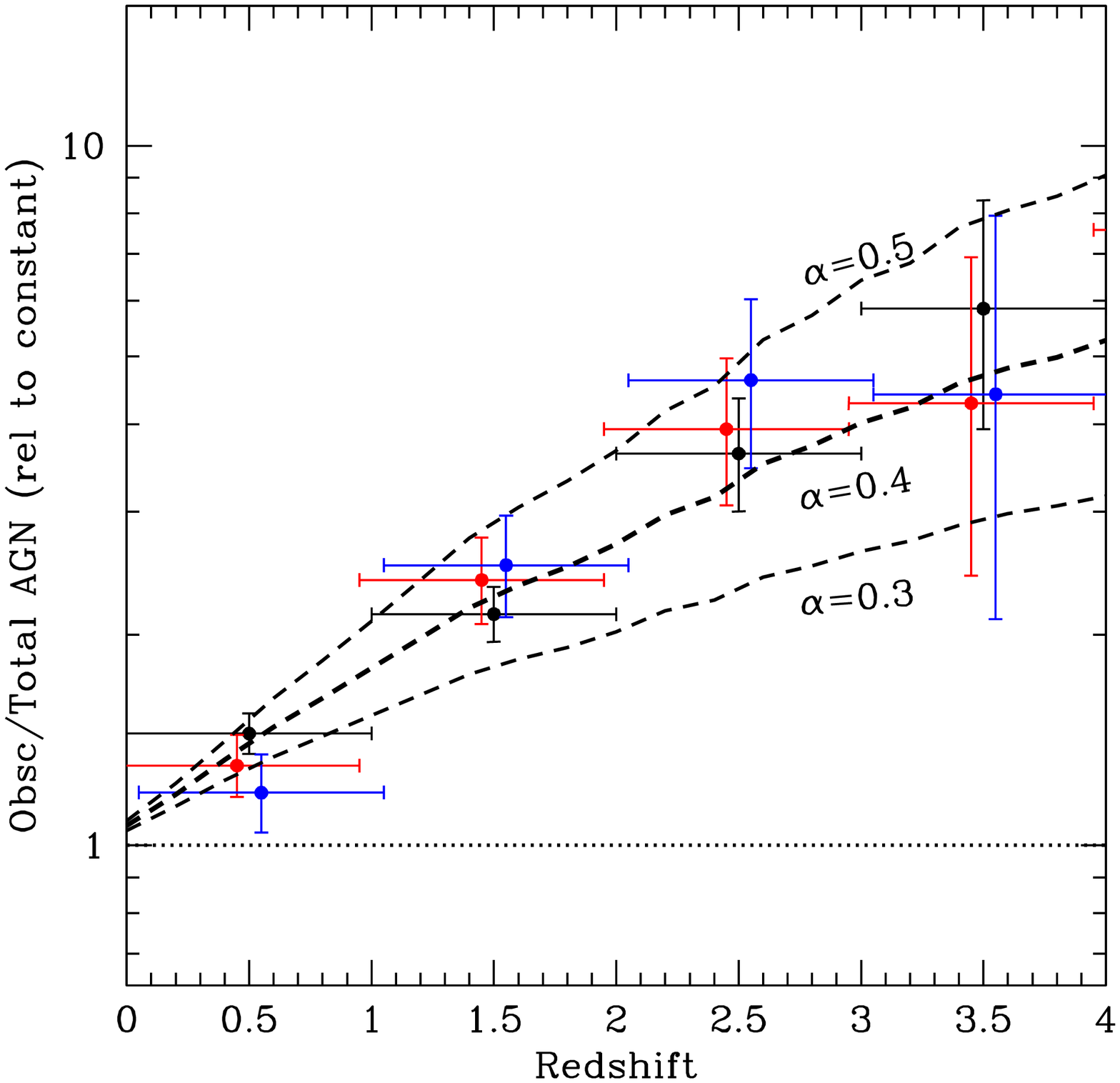}
\caption{({\it Left}) Observed fraction of obscured AGN 
as a function of redshift. 
({\it Data points}) Measured fraction for our super-sample of 
1229 optically-classified AGN {(\it black points}),
and for sub-samples cut at brighter optical magnitudes 
to increase the spectroscopic completeness 
to 73\% {(\it red points}) and 93\% {(\it blue points}). 
({\it Solid lines}) Expected fraction as a function of redshift 
for an intrinsically non-evolving ratio, 
taking into account the effects of spectroscopic incompleteness 
and flux limits for the super-sample and two sub-samples.
({\it Right}) Fraction of obscured AGN relative to 
the expectations for a non-evolving obscured AGN ratio ({\it dotted line}), 
incorporating the effects of spectroscopic incompleteness. 
A significant increase with redshift is clearly seen, 
independent of the spectroscopic completeness of the sample. 
{\it Dashed lines} show an intrinsic evolution of the form (1+$z$)$^{\alpha}$, 
for $\alpha=0.4 \pm 0.1$.
}
\label{RatioEvol}
\end{center}
\end{figure}

Figure~\ref{RatioEvol}a shows the observed fraction 
as a function of redshift ({\it black data points}).
As many have noted previously, the observed fraction declines with redshift,
from roughly 3/4 locally to $\sim1/3$ at $z\sim4$.
This has been interpreted to mean that obscured AGN are rare at high redshift.
However, the lines in Figure~\ref{RatioEvol}a 
show the expected decline for an underlying population 
whose obscured fraction is actually constant with redshift,
calculated for our super-sample using the appropriate corrections
for X-ray and optical flux limits and spectroscopic completeness.
That is, an even steeper decline is expected in the observed samples
even when the underlying population does not change at all with redshift.

Three significant selection effects cause this strong decline.
First, obscured AGN have smaller X-ray fluxes, so there is a bias
against their detection in the X-ray sample in the first place.
This is a small effect, particularly because it becomes relatively
less important with increasing redshift 
(for which rest-frame emission is in an increasingly 
harder X-ray band affected only by higher and higher column densities).
Second, obscured AGN are fainter in the optical, so there is a bias
against spectroscopic identifications for $z\simgt 1$ \citep{treister04};
this is not important for samples with highly complete
spectroscopic identifications.
Third and quite important, the obscured fraction depends strongly
on luminosity (Fig.~\ref{figLumRatio}). 
Given the luminosity-redshift correlation inherent in flux-limited samples,
the mean AGN luminosity increases with redshift and therefore
the obscured fraction that is observed decreases --- even if the same population of
obscured lower-luminosity AGN is present.
Our analysis shows that this is the dominant selection effect.
It is an important effect even in AGN samples with 100\% complete
spectroscopic identifications, as indicated by the blue line in 
Figure~\ref{RatioEvol}a.
Simply put, high redshift AGN samples are biased to higher luminosity, 
and thus contain lower obscured fractions, 
even if there is no underlying evolution of the ratio between obscured
and unobscured AGN at all.

The observed more gradual decline in obscured fraction of AGN
actually implies an increase in obscured fraction with redshift. 
Figure~\ref{RatioEvol}b shows the data relative to 
(i.e., divided by) the expectation for a non-evolving population, 
for the super-sample and the two sub-samples 
with higher identification fractions.
The effect is largely independent of the 
completeness of the optical identifications.
Fitting the increasing fraction with a simple power-law dependence
on redshift, $\propto (1+z)^\alpha$, gives a good fit for 
$\alpha = 0.4 \pm 0.1$.
This means that the fraction of AGN at redshift $z\sim4$ 
that are obscured is observed to be twice as high as would
be the case were the intrinsic fraction constant.
AGN obscuration is substantially greater in the young Universe.

\section{Summary}

GOODS, MUSYC, and other deep multiwavelength surveys provide
overwhelming evidence for a large population of obscured AGN
that dominate AGN demographics out to high redshifts.
Optical surveys are biased against detecting these objects, 
and even hard X-ray surveys, which are considerably
less biased, suffer strong selection effects,
primarily due to the luminosity-dependence of obscuration. 

Taking these effects into account quantitatively, 
with a realistic, well-constrained population synthesis model 
that uses AGN spectral energy distributions based on a unification paradigm,
we deduce that the ratio of obscured ($N_H > 10^{22}$~cm$^{-2}$)
to unobscured ($N_H < 10^{22}$~cm$^{-2}$) AGN is roughly 3:1 locally
(integrated over all luminosities), and increases with redshift. 
Low-luminosity AGN ($10^{42}$~ergs/s $< L_X < 10^{44}$~ergs/s) are
much more likely to be obscured than high-luminosity AGN
($L_X > 10^{44}$~ergs/s).

To the extent that our assumed infrared through X-ray 
spectral energy distributions are reasonable, 
and our assumed $N_H$ distribution is reasonable 
(it is essentially the same as that used by others in the field; 
see \citealp{gilli07} for a comparison of the different distributions), 
these results are completely robust.

How might one avoid the biases inherent in optical and X-ray surveys?
In principle, far-infrared surveys are unbiased
because the absorbed energy is re-radiated thermally;
however, infrared surveys are very inefficient for AGN selection
since the infrared sky is strongly dominated by starlight (Fig.~\ref{figIRBG}a).
In addition, AGN signatures (such as broad emission lines, 
strong power-law continua, and rapid variability) 
may well be hidden in obscured objects.
Thus, identifying complete samples of AGN from far-infrared surveys 
will never be a simple matter, although it can potentially
put useful limits on the fraction of galaxies with buried AGN.

Selection effects are very important to take into account for any
survey, even those that are 100\% identified. 
For example, consider a deep hard X-ray survey for which all sources 
have optical and/or infrared counterparts with known redshifts and
classification.
Many would assume the survey itself is not strongly biased and
that the identifications yield the underlying demographics, 
since no X-ray sources remain unidentified.
However, missing from the X-ray sample are the most heavily obscured AGN; 
even more important, unobscured AGN are over-represented
(relative to their fraction of the underlying population), 
especially at high redshift,
because of the dependence of obscuration on luminosity.
This in fact is the dominant effect for existing surveys.

Therefore, to understand the distribution of black holes in the universe
and to estimate cosmic accretion rates, 
the selection effects must be modeled using reasonable
assumptions about the underlying population.
That in turns yields a picture of the universe that matches well
our picture of nascent accreting black holes at the centers of
dusty, star-forming, young galaxies in the early Universe.

\acknowledgments
This work would not have been possible without the exquisite data
made possible by NASA's Great Observatories and by 
great observatories on the ground.
We gratefully acknowledge support from
NSF grant AST-0407295      
and NASA grants 
NAG5-10301, 
HST-AR-10689.01-A, 
HST-GO-09822.09-A, 
HST-GO-09425.13-A, 
and NNG05GM79G. 



%

\end{document}